\documentclass[preprint,12pt,groupaddress]{revtex4-1}

\usepackage{color}
\usepackage{lipsum}
\usepackage{float}
\usepackage[titletoc,toc,title]{appendix}
\usepackage{amssymb}
\usepackage{epsfig}
\usepackage{amsmath}
\usepackage{subfigure}
\usepackage{graphicx}
\usepackage{booktabs}
\usepackage{colortbl}
\usepackage{natbib}
\usepackage{color}
\usepackage{bbm}

\newcommand{\nobracket}{}
\newcommand{\nocomma}{}
\newcommand{\noplus}{}

\newcommand{\tmem}[1]{{\em #1\/}}
\newcommand{\tmmathbf}[1]{\ensuremath{\boldsymbol{#1}}}
\newcommand{\tmop}[1]{\ensuremath{\operatorname{#1}}}
\newcommand{\tmstrong}[1]{\textbf{#1}}
\begin{document}

\title{Dynamics in fractal spaces}

\author{\'{A}lvaro G. L\'{o}pez}

\address{Nonlinear Dynamics, Chaos and Complex Systems Group.\\Departamento de F\'isica, Universidad Rey Juan Carlos, Tulip\'an s/n, 28933 M\'ostoles, Madrid, Spain}

\date{\today}

\begin{abstract}
We study the dynamics of a particle in a space that is non-differentiable. Non-smooth geometrical objects have an inherently probabilistic nature and, consequently, introduce stochasticity in the motion of a body that lives in their realm. We use the mathematical concept of fiber bundle to characterize the multivalued nature of geodesic trajectories going through a point that is non-differentiable. Then, we generalize our concepts to everywhere non-smooth structures. The resulting theoretical framework can be considered a hybridization of the theory of surfaces and the theory of stochastic processes. We keep the concepts as general as possible, in order to allow for the introduction of other fundamental processes capable of modeling the fractality or the fluctuations of any conceivable continuous, but non-differentiable space.\\
\end{abstract}

\maketitle
%\pacs{05.45.Ac,05.45.Df,05.45.Pq}

%\keywords{Dynamics - Fractals - Stochastic processes - Non-differentiable geometry}

%%%%%%%%%%%%%%%%%%%%%%%%%%%%%%%%

\section{Introduction}

It has been suggested in the last decades that quantum behavior might be a
consequence of the fractal nature of spacetime \cite{nott1,nott2}. As it has
been recently demonstrated, microscopic charged bodies can self-oscillate,
which endows their surrounding space with rapidly fluctuating electromagnetic
fields \cite{lope,lop202}. These fields can be interpreted, according to the theory of general relativity, as a very pronounced curvature of spacetime. In this sense, the hypothesis of a fractal space should be very welcomed among scientists and deserves more rigor and attention. Although unnecessary to explain quantum behavior from a strict point of view, this new perspective might well serve to model the dynamics of charged particles experiencing not only zitterbewegung but strong and fast fluctuations of any nature, which take place in a region of space where the quantum potential emerges from a very complicated interaction of many bodies. For example, in an electron gas, it has been shown that a description in terms of branching flows can be admitted \cite{topi}.

The theory of fractal spacetime has been built upon the theory of conservative diffusion \cite{nels1} developed by Edward Nelson to show that a kinematic theory of quantum mechanics is feasible \cite{nels2}.
Stochastic mechanics is, from the mathematical point of view, totally
equivalent to non-relativistic quantum mechanics. It is based on the idea that two stochastic diffusive processes, one going forward in time and another going backward in time, can be superposed to yield a process of conservative diffusion \cite{bacc}. When built by means of the Wiener process, whether using Ito's or Stratonovich's calculus frame, the two stochastic differential
equations are fundamentally irreversible. Moreover, the drifts are not \emph{a priori} the same, and a twofoldness appears naturally from the very beginning of the theory. As it has been demonstrated by Nelson, the reversed process can be related to the derivative when variations are taken from the left. Since stochastic processes are non-differentiable, the limit of a function when a point is approached from the left differs from the limit as taken from the right. Then, if the forward process leads to a diffusive Fokker-Planck equation governing the time evolution of the probability density, the reverse process is guided by an antidiffusive Fokker-Planck equation. The superposition of these two processes leads to an equation of probability conservation, just as it appears in non-relativistic quantum
mechanics.

This state of things has been cleverly used by Nottale to develop a theory of fractal spacetime \cite{nott1}. The twofoldness appearing in stochastic mechanics is used to create complex kinematic variables, which lead to the definition of a complex action, from which the Schr\"odinger equation can be readily derived. Nevertheless, the ambition of this author extends beyond first quantization and quantum dynamics. Since fractal structures have resolution at all scales, the author tries to extend the principle of covariance to scale transformations \cite{nott2} and proposes a relativistic law of scale multiplication, similar to Einstein's law for the addition of velocities in special relativity. 

Since the introduction of stochastic mechanics by Nelson, other pioneering works have dedicated their efforts to the study of motion in complicated physical spaces \cite{fred,prug,kuri}. Some of these approaches are based on the concept of stochastic space and frequently introduce a probability density depending on the coordinates of space to represent the chance that a point mass is observed at different phase space locations \cite{prug,kuri}. However, since most of these studies have been developed with the intention to give a geometrical foundation to the quantum formalism, they frequently rely on a set of additional postulates and principles. For example, in Frederick's seminal work \cite{fred} five postulates are considered, among which a metric superposition condition is assumed. In a more recent work \cite{kuri}, an extremal entropy principle is introduced in order to reconcile the stochastic dynamics with path integral quantization. As far as we are concerned, none of these works derive their dynamical equations from purely classical geometrical considerations and at the same time cover in a simple way the theory of smooth manifolds as a limiting physical case, when fluctuations are negligible.

In the present work we pose the problem of motion in a fractal space from a more elementary and geometrical point of view. This new perspective has the advantage of providing covariant stochastic differential equations and including the typical geodesic motion that takes place in smooth manifolds in the limit of negligible fluctuations. It is not our purpose to derive quantum mechanics from our approach, even though the present framework can be connected to Nelson's stochastic mechanics by considering appropriate limits. For simplicity, but without loss of generality, we consider the specific problem of bodies moving in two dimensional rough surfaces embedded in three-dimensional space and assume conventional diffusion processes to model fluctuations. We first study the dynamics of a point particle on a manifold with a single non-differentiable point, showing that non-differentiability introduces randomness in a simple and logical way. Then, we consider a space where a denumerable set of points that are not smooth can be found, obtaining a random walk. Finally, a continuum limit is worked out, which leads naturally to stochastic dynamics. We derive Langevin covariant equations describing the particle's dynamics and compute their corresponding Fokker-Plank equations governing the evolution of the probability density for two specific examples. Several possible applications of the present mathematical framework are pointed out in the closing section.

\section{Mechanics in non-differentiable Riemannian spaces}

In order to understand the mechanics of a particle in a non-smooth Riemmanian
space, we proceed constructively, analyzing the dynamics of a particle at a
non-differentiable point in the first place. We start from first principles
and make use of standard tools used in classical differential geometry
\cite{krey}. We introduce no other logical assumptions, except for the
principle of indifference of Laplace \cite{lemo}, to characterize
the degeneracy of coexisting tangent spaces at a non-differentiable point.
Once this problem is properly solved, the extension of these ideas to more
complicated spaces is rather simple by induction and, finally, to fractal
spaces by taking a limit to the continuum.

\subsection{A manifold with one non-differentiable point}

\begin{figure}
\centering
  \resizebox{300pt}{220pt}{\includegraphics{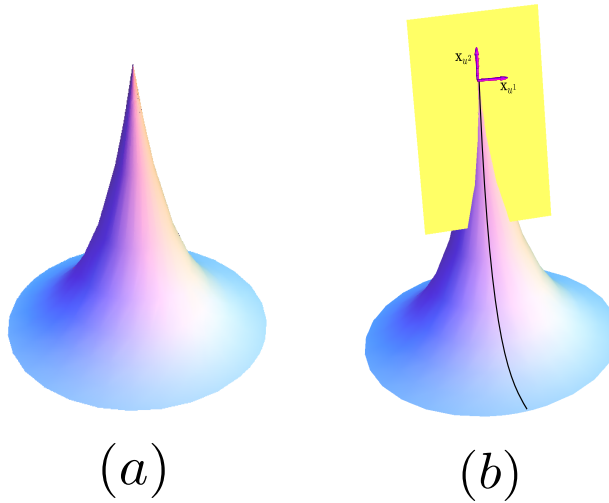}}
  \caption{{\tmstrong{A conical point.}} (a) A surface with a
  non-differentiable point. (b) A limiting tangent space (yellow) when the conical
  point is approached through a geodesic (black line).}
  \label{fig:1}
\end{figure}

We begin our study with the most simple case of a \emph{non-differentiable
space}. For this purpose, we consider a two-dimensional manifold $\mathcal{M}$ embedded in the
euclidean space $\mathbbm{R}^{3}$ with a single non-differentiable point, as for example the one given by the coordinate system $(u_1, u_2)$ through the parametrization
\begin{equation} 
\tmmathbf{x} (u^1, u^2) = \left( u^1, u^2, e^{- \sqrt{(u^1)^2 + (u^2)^2}}
   \right),
\end{equation}
with $u^1, u^2 \in \mathbbm{R}$. The resulting surface is shown in Fig.~\ref{fig:1} and
it locally resembles a cone. The point $P (0, 0, 1)$ is clearly
non-differentiable, since an infinite set of tangent planes coexist at it. By
means of this parametrization, we can compute the tangent vectors at $P$ if we
approach the point appropriately. For example, suppose that we approach a point
along the curve $(0, t)$, which is a geodesic of the manifold. Then, the
tangent space is spanned by the two vectors
\begin{equation} 
\tmmathbf{x}_{u^1} = 
     \lim_{t \rightarrow 0^+} \left( \dfrac{\partial \tmmathbf{x}}{\partial
     u^1} \right)_{(0, t)} = (1, 0, 0) ,~~\tmmathbf{x}_{u^2} = \lim_{t
     \rightarrow 0^+} \left( \dfrac{\partial \tmmathbf{x}}{\partial u^2}
     \right)_{(0, t)} = (0, 1, - 1).
\end{equation}

On the other hand, consider that we approach the point along $(t, 0)$, which
is another geodesic, orthogonal to the previous one. Now, we get another
tangent space, generated by the two linearly independent vectors
\begin{equation} 
\tmmathbf{x}_{u^1} = 
     \lim_{t \rightarrow 0^+} \left( \dfrac{\partial \tmmathbf{x}}{\partial
     u^1} \right)_{(t, 0)} = (1, 0, - 1) ,~~ \tmmathbf{x}_{u^2} = \lim_{t
     \rightarrow 0^+} \left( \dfrac{\partial \tmmathbf{x}}{\partial u^2}
     \right)_{(t, 0)} = (0, 1, 0).
\end{equation}

In these two cases, the couple of limiting vectors are
orthogonal and they span the whole tangent space. More generally, we can consider the limit
\begin{equation} 
     \tmmathbf{x}_{u^1} (\alpha) = \lim_{t \rightarrow 0^+} \left(
     \frac{\partial \tmmathbf{x}}{\partial u^1} \right)_{(t \cos \alpha, t
     \sin \alpha)} = (1, 0, - \cos \alpha),
\end{equation}
for the first vector, and the corresponding limit
\begin{equation} 
     \tmmathbf{x}_{u^2} (\alpha) = \lim_{t \rightarrow 0^+} \left(
     \frac{\partial \tmmathbf{x}}{\partial u^2} \right)_{(t \cos \alpha, t
     \sin \alpha)} = (0, 1, - \sin \alpha),
\end{equation}
for the second. We note that, in all of the previous examples, we have approached the
point along geodesic curves of the surface, which emanate from the
non-differentiable point $P (0, 0, 1)$. Thus, given a geodesic curve that is
born at the point $P$, we can construct a tangent plane along the direction
specified by such geodesic curve. Hereafter, we shall define any point of this kind as a \emph{conical point}. As we have demonstrated, at such
a point there exist an infinite number of tangent spaces $T_p^{(c^1, c^2)}
\mathcal{M}$, each of them generated by approaching the point along a
particular geodesic direction $(c^1, c^2)$.

Instead of disregarding this coexistence of tangent spaces as something
pathological or undesired, we include all of them in one set $T_p \mathcal{M}$
and assume that the \emph{tangent space} at such point is
\emph{multivalued}. Mathematically, this requires to consider the
union of all the tangent spaces
\begin{equation} 
T_p \mathcal{M}= \bigcup_{(c^1, c^2) \in S^1} T^{(c^1, c^2)}_p
   \mathcal{M}, 
\end{equation}
where $(c^1, c^2)$ denote the tangent space when the conical point is
approached in a particular direction. As an example, in the case previously addressed we can consider the parametrization $(c^1 (\alpha), c^2 (\alpha)) =
(\cos \alpha, \sin \alpha)$, so that the bundle is written explicitly as
\begin{equation} 
T_p \mathcal{M}= \{ (x, y, z, \alpha) \in \mathbbm{R}^3 \times S^1 |
   \nobracket x \cos \alpha + y \sin \alpha + z = 1 \}.
\end{equation}    

Now, let us think of a particle moving in a manifold of this nature. Assume that the
particle starts its journey at some point and follows a geodesic towards the
apex of the cone. When such point is reached, since there exist an infinite
number of tangent spaces at it, and there is no reason for the particle to
prefer one of them at the expense of the others, from a logical point of view,
we see ourselves forced to admit that any of them can be followed. This is
nothing else than the principle of indifference, formulated a long time ago by
Laplace \cite{lapl}. The dynamics at this point thus becomes \tmem{inherently random}, and a probability distribution has to be attached to any point of this nature to choose among one of the possible tangent spaces when a particle traverses it.

From the point of view of the theory of differential equations, we might say
that at a non-differentiable point, the \tmem{uniqueness} of the solutions is
lost, and the dynamics becomes probabilistic in an essential way. However, and
as we have shown, there \tmem{exist} solutions for the different geodesics
that meet at the conical point, since the limit in a certain direction is
well-defined. In particular, given a conical point, there exists an infinite
set of geodesics emanating from it. Therefore, it seems natural to use
geodesic coordinate systems to characterize the dynamics of the particle when
it arrives at an isolated conical point.

Given a smooth manifold $(u^1, u^2)$ and a sufficiently close neighborhood
$U$ of some point $P (u^1_0, u^2_0)$ in it, it is always possible to construct
a coordinate system from the geodesics that transverse it (\tmem{i.e.} from
a geodesic field in $U / \{ P \}$). For this purpose, consider the geodesic
equations of the manifold, which are described by the differential equations
\begin{equation} 
\ddot{u}^i + \Gamma^i_{j k} \dot{u}^j \dot{u}^k = 0,
\end{equation}
with initial conditions
\begin{equation} 
u^i (0) = u^i_0, \dot{u}^i (0) = c^i,
\end{equation}
where the derivative represented by the dots must be considered with respect
to the arclength and the Christoffel symbols $\Gamma^i_{j k}$ have been introduced. We can represent these solutions in the form
\begin{equation} 
u^i = \Phi^i (s, c^1, c^2).
\end{equation} 

\begin{figure}
\centering
  \resizebox{300pt}{180pt}{\includegraphics{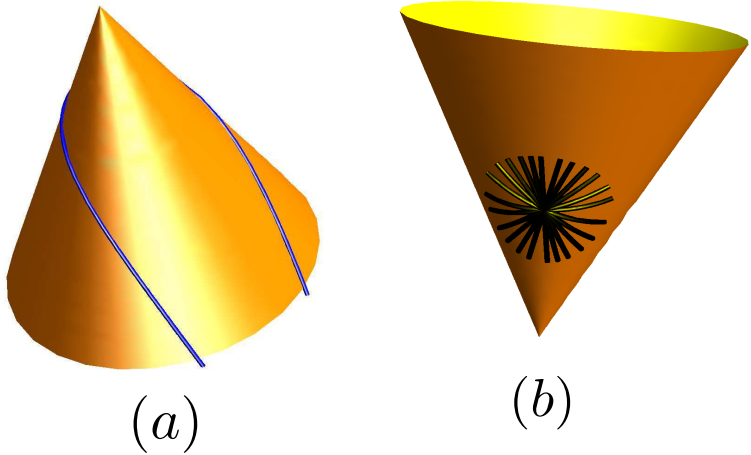}}
  \caption{\tmstrong{Riemann normal coordinates.} (a) A geodesic path going
  around a cone. (b) A set of geodesics emanating from a point, which allow to
  define the radial part of a polar coordinate system at such point.}
\label{fig:2}
\end{figure}

From this coordinate system, another coordinate system $\{ v^i \}$ can be
designed, in such a manner that, in the new coordinates, the motion is
uniform. This new coordinate system is simply related to the previous
one by means of the exponential map $\exp: T_{p} \mathcal{M} \rightarrow \mathcal{M}$, which allows to lower elements from the tangent space to the manifold. This map is defined as
\begin{equation}
u^i = \Phi^i (1, v^1, v^2).
\end{equation} 
If the point is a smooth one, in such a coordinate system the space can be
regarded as a locally flat space and the metric can therefore be defined to be
the identity. These coordinates are commonly known as \tmem{Riemann
coordinates}, and the geodesic equations emanating from the point can be
parametrized in a particularly simple manner $v_i (s) = c_i s$ by means of them. In other words, the motion is uniform in the new coordinate system. As it is well known, these coordinate system corresponds to free falling inertial observers in relativistic theory \cite{eins}, where gravitational effects locally disappear. To familiarize the reader with such coordinate systems, in Appendix A we provide a simple example by computing the Riemann coordinates of a smooth point belonging to the slope of a cone, as shown in Fig.~\ref{fig:2}.

This coordinate system can be constructed in the neighborhood of a conical
point as well because, even though a unique limit does not exist at the
point, and in the traditional sense there is no limit, given a certain
direction, it does certainly exist. We insist further on this idea, because it is of fundamental importance to the development of present work. The
construction of the coordinate system is possible because, starting at a conical point, even though the tangent vectors might not be uniquely defined at the point itself, there is one geodesic emanating from it in each direction and, therefore, given a geodesic curve, the tangent vector can be perfectly computed, as we have thoroughly done in the previous paragraphs.

Consequently, we can assume that when a particle reaches a conical point through one of the geodesics that sink there, the Riemann coordinates defined in its neighborhood experience a random linear transformation $L$, and the particle continues its motion through some other geodesic. Thus, the following linear transformation between two of the tangent spaces that coexist at the conical point is naturally defined
\begin{equation} 
L : T_p^{(c^1, c^2)} \mathcal{M} \rightarrow T_p^{(\tilde{c}^1,
   \tilde{c}^2)} \mathcal{M},
\end{equation} 
where $\tilde{c}^i = L^i_j c^j$. This transformation simply switches between two tangent spaces belonging to the bundle $T_p \mathcal{M}$ previously
introduced. In other words, $L$ interchanges the dynamics from one of the
geodesics that traverses the conical point to another. Consequently, the
transformation $L$ induces another Riemann coordinate system $\{ \tilde{v}^i
\}$, which is obviously related to the previous one through the transformation
\begin{equation} 
\tilde{v}^i = L^i_j v^j.
\end{equation}

The new geodesic should be chosen at random, according to some probability
distribution $p (L_j^i)$. Thus, the matrix $L$ representing the transformation
at the conical point is a \tmem{random matrix}, consisting of four random
variables $\{L^i_j \}$. As will be cleared in the following section, if energy is to be conserved, we must require to this matrix to be an orthogonal one, $i.e.,$ we have to impose $L \in \tmop{SO} (2, \mathbbm{R})$. We name
this matrix as $R$ hereafter, not to forget that it represents a rotation. In Appendix B we construct this rotation for a particle traversing the apex of a simple cone, to manifestly see how these two coordinate systems are related.

We can write the coordinate transformations connecting a particular coordinate
system $u^i$ to the Riemann coordinates $v^i$ in the form
\begin{equation} 
u^i = \Phi^i (v^1, v^2),
\end{equation}
and consequently $v^i = \Psi^{i} (u^1, u^2)$, where $\Psi$ represents the inverse mapping of $\Phi$. By applying a linear transformation at the conical point, we switch to another Riemann coordinate system on the manifold $\tilde{v}^i$, to which another coordinate system $\tilde{u}^i$ is associated. Having assumed an arrow of time, we adopt this notation henceforth, where the tilde stands for the coordinate system of the particle exiting the cone, while bare variables correspond to the incoming particle. Now, we have
\begin{equation} 
\tilde{u}^i = \Phi^i (\tilde{v}^1, \tilde{v}^2) \Rightarrow \tilde{v}^i =
   \Psi^i (\tilde{u}^1, \tilde{u}^2).
\end{equation}

A tangent vector at the conical point $\dot{v}^i$ is transformed into a new
tangent vector $\dot{\tilde{v}}^i$, according to the relation
\begin{equation} 
\dot{\tilde{v}}^i = R^i_j \dot{v}^j.
\end{equation} 

On the other hand, we have that the Riemann coordinate system is related to
the original one through the relation
\begin{equation}
\dot{u}^i = \frac{\partial \Phi^i}{\partial v^j}  \dot{v}^j = J^i_j
   \dot{v}^j,
\end{equation}
where $J^i_j$ is the Jacobian matrix between the Riemann coordinates and the
original ones, which again is adequately defined in the limiting sense
previously discussed. Concerning the new coordinate system to which the
particle switches when going through the conical point, we have
\begin{equation}
\dot{\tilde{u}}^i = \dfrac{\partial \Phi^i}{\partial \tilde{v}^j}
   \dot{\tilde{v}}^j = \dfrac{\partial \Phi^i}{\partial \tilde{v}^j} R^j_k
   \dot{v}^k = \dfrac{\partial \Phi^i}{\partial \tilde{v}^j} R^j_k
   \dfrac{\partial \Psi^k}{\partial u^l} \dot{u}^l,
\end{equation}
which can be more neatly written as
\begin{equation}
\begin{array}{l}
     \dot{\tilde{u}}^i = \tilde{J}^i_j R^j_k (J^{- 1})^k_l \dot{u}^l.
   \end{array}
\end{equation}

\begin{figure}
\centering
  \resizebox{120pt}{80px}{\includegraphics{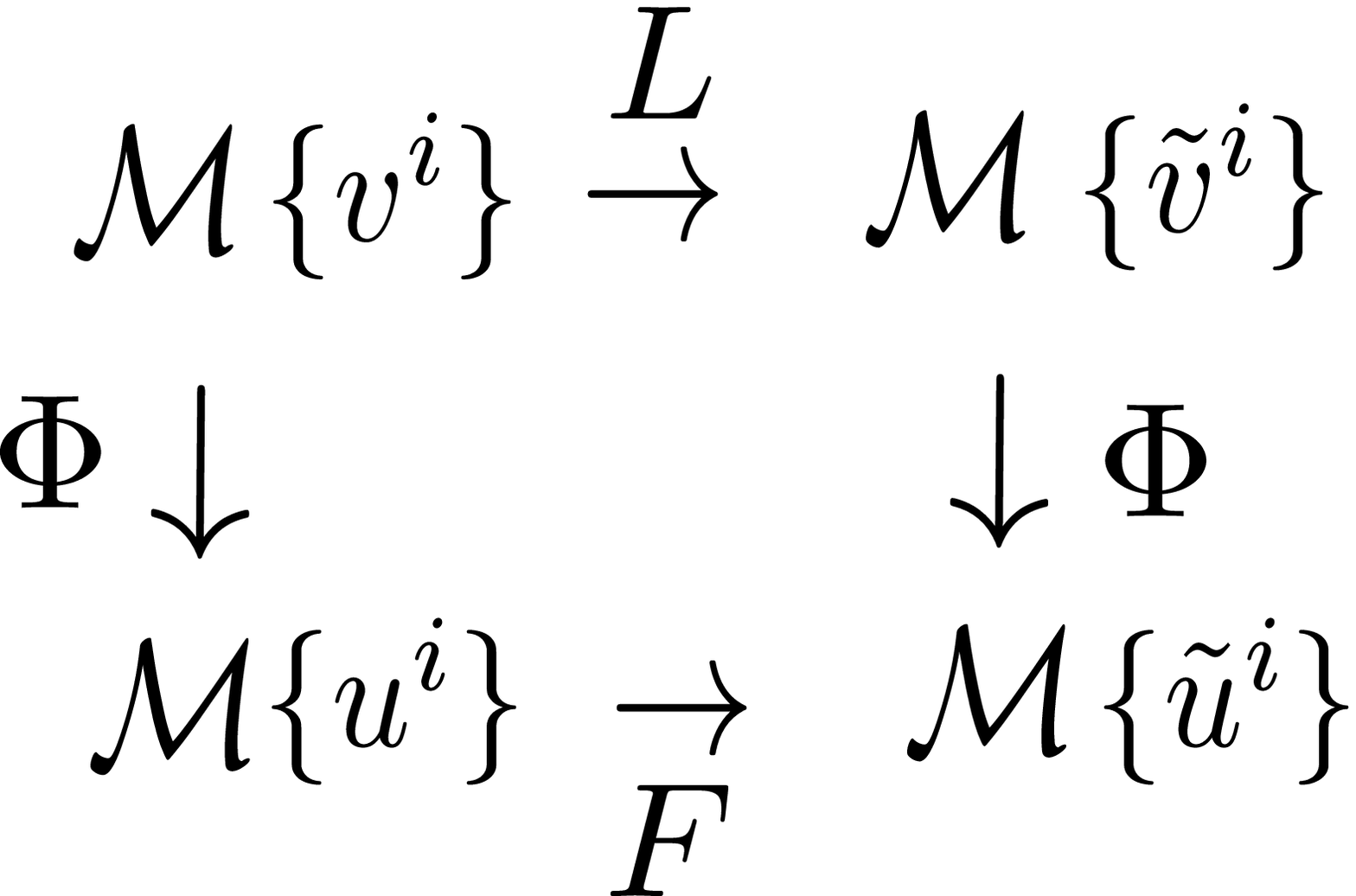}}
  \caption{\textbf{Switching coordinate systems}. A commutative diagram showing the relations among normal coordinates and some ordinary coordinate system at a conical point. The linear random mapping $L$ between the incoming and the outgoing Riemann coordinate systems induces a nonlinear transformation between the coordinates at the entrance and at the exit of the conical point.}
  \label{fig:3}
\end{figure}

In fact, given the commutative diagram shown in Fig.~\ref{fig:3}, we must have a coordinate transformation $F$ between the initial coordinate system and a new
one in the form
\begin{equation}
\tilde{u}^i = F^i (u^1, u^2),
\end{equation}
where $F = \Phi \circ R \circ \Psi$ and
\begin{equation}
\dot{\tilde{u}}^i = \frac{\partial F^i}{\partial u^j}  \dot{u}^j.
\end{equation}
Since $R$ belongs to a group and $J$ is also a linear transformation, $F$ can
be regarded as an adjoint mapping. Concerning the basis, we have that
\begin{equation}
\dot{\tmmathbf{x}} = \dfrac{\partial \tmmathbf{x}}{\partial u^i} \dot{u}^i =
\dot{u}^i \tmmathbf{x}_{u^i} = \dot{v}^j \dfrac{\partial \Phi^i}{\partial
v^j} \tmmathbf{x}_{u^i},
\end{equation}
what imposes the relation between tangent basis $\tmmathbf{x}_{v^j} = J^i_j
\tmmathbf{x}_{u^i} $. But more importantly, we have also that the tangent vectors experience a random change when they go through the conical point
\begin{equation} 
\dot{\tmmathbf{x}} = \dot{\tilde{u}}^i \tmmathbf{x}_{\tilde{u}^i} =
   \tilde{J}^i_j R^j_k (J^{- 1})^k_l \dot{u}^l \tmmathbf{x}_{\tilde{u}^i},
   \label{eq:23}
\end{equation}
which now leads to the identity $\tmmathbf{x}_{u^l} = \tilde{J}^i_j R^j_k (J^{- 1})^k_l\tmmathbf{x}_{\tilde{u}^i}$. Thus we have demonstrated that the dynamics at the non-differentiable point can be described as a change between two coordinate systems, randomly executed. In short, Eq.~\eqref{eq:23} proves that for a particle moving on a non-smooth space, when it arrives at a conical point, the tangent vector experiences a random linear transformation that switches from one of the geodesics that is born at such point, to one of its sister geodesics. This switch between geodesics can ultimately be regarded as a change of coordinates.

From a mathematical point of view, we can formalize these ideas by introducing a fiber bundle \cite{naka} to represent a space that is
non-differentiable at some point. For this purpose, we consider a Riemann
polar coordinate system (homeomorphic to $S_1 \times \mathbbm{R}_{> 0}$) at
the conical point and take the $S_1$ part of these coordinate system as the
base of a fiber bundle, where the fibers are the tangent vectors of $T^{(c_1, c_2)} M$, computed as previously described. Then, the manifold $S_1$ can be covered with the two typical charts $(U_1, \varphi_1)$ and $(U_2, \varphi_2)$, where $U_1 = (0, 2 \pi)$, $U_2 = (- \pi, \pi)$, and $\varphi_1$ and $\varphi_2$ are two smooth coordinate maps from the base space to $\mathbbm{R}$. These two charts are connected between them by means of a rotation. In fact, a third chart is necessary to combine points outside $U_1 \cap U_2$, but we omit it for simplicity. Therefore, the transition function of the fiber bundle $R$ is a transformation in the form $R : U_1 \cap U_2 \rightarrow \tmop{SO} (2)$, where $R (\theta)$ is a rotation. However, the procedure for connecting the two fibers $T^{(c_1, c_2)}\mathcal{M}$ and $T^{(\tilde{c}_1, \tilde{c}_2)}\mathcal{M}$ of the bundle is more sophisticated than the traditional examples found in the study of differentiable manifolds, insofar as the matrix entries of $R$ are chosen at random. Consequently, a probability distribution together with the structure group of the bundle has to be provided to describe the dynamics in a non-smooth space. Simply put, the fibers are now randomly intertwined which, to some extent, gives rise to a geometrical structure that resembles the nest of a bird. 

In conclusion, we can rigorously describe the dynamics at a conical point utilizing a stochastic dynamical fiber bundle. Nevertheless, we shall not dwell on these mathematical technicalities any longer, to concentrate our following efforts on the study of the dynamics in a manifold where many non-differentiable points are present. As the reader may have already guessed, the existence of numerous non-differentiable points gives rise to a random walk dynamics along the space.

\subsection{A space with an infinite denumerable set of non-differentiable
points}

\begin{figure}
\centering
  \resizebox{380pt}{220pt}{\includegraphics{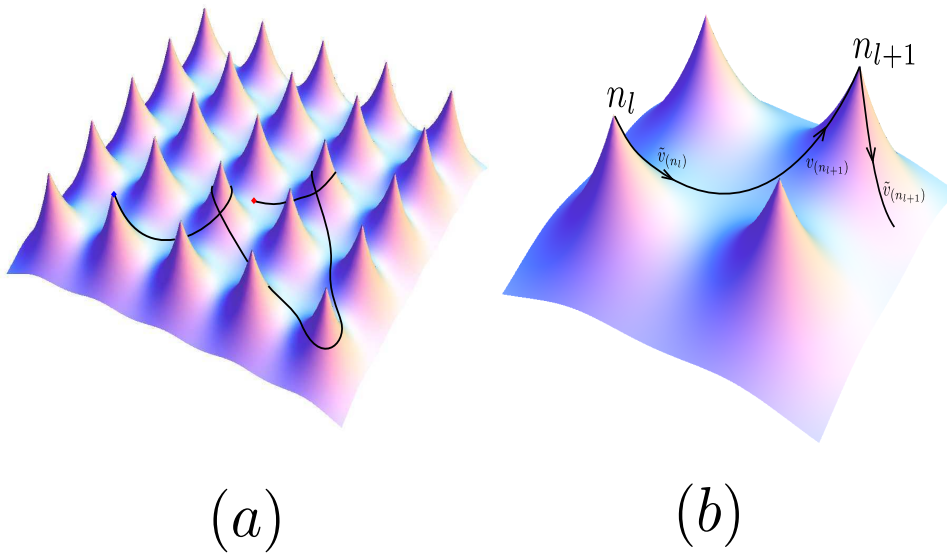}}
  \caption{\tmstrong{A non-differentiable space.} (a) A space with many conical points distributed all over it. A geodesic path moving chaotically through this complicated space. (b) Several Riemann coordinate systems appear in sequence connecting different conical points.}
  \label{fig:4}
\end{figure}

We increase the complexity of the space by allowing at most an infinite
denumerable number of non-differentiable points, all of them at a finite distance (from below) from each other. In other words, the conical points are not allowed to fill densely the manifold, so that the distance between them can not be infinitesimal. A subset illustrating part of such a manifold is depicted in Fig.~\ref{fig:4}. We now enumerate the conical points using the index $n$. For any two conical points, there exists a sequence of geodesics that communicates them. As shown in the previous section, we can attach a set of Riemann coordinates $v_{(n)}^i$ to each conical point $n$ and, for simplicity and without loss of generality, suppose that the dynamics is constrained to the geodesics connecting these points. In practice, if the dynamics is not restricted to these points, a chaotic type of movement can arise in the present situation, which is very similar to a random walk, but deterministic (see Fig.~\ref{fig:4}(a)). However, interesting as it may be, we will not indulge ourselves with this problem, since our ultimate purpose is to study the dynamics of a body living in a nowhere differentiable manifold, which is the subject of the forthcoming sections.

With this restriction in mind, it is evident that the particle moves between
conical points following geodesics, describing a \tmem{random walk} through
the lattice of conical points. Just for clarity in the exposition, and because geodesic trajectories are always traveled at a constant speed, we choose a description in terms of time instead of arclength henceforth. More specifically, we have a coordinate system in each segment $l$ of the
random walk for times $t_l < t < t_{l + 1}$ given by the geodesic equations
\begin{equation}
\tilde{u}_{(n_l)}^i (t) = \Phi^i_{(n_l)} (t - t_l, \tilde{c}^1_{(n_l)},
   \tilde{c}^2_{(n_l)})_{}, 
\end{equation}
with initial conditions
\begin{equation}
\tilde{u}^i (t_l) = \tilde{u}^i_l, \dot{\tilde{u}}^i (t_l) =
   \tilde{c}^i_{(n_l)}.
\end{equation}

Following previous definitions, to each conical point $(n_l)$ visited at time $t_l$, we attach two sets
of Riemann coordinates, one representing the particle entering the point
$v_{(n_l)}^i$ and another getting out of it $\tilde{v}^i_{(n_l)}$, after
experiencing a linear transformation, as specified in the previous section.
Accordingly, the relations
\begin{equation} 
\tilde{c}_{(n_l)}^i = R^i_j (l) c^j_{(n_l)},  
   \label{eq:26}
\end{equation}
and $\tilde{v}_{(n_l)}^i = R^i_j (l) v_{(n_l)}^j$ hold. Note that we have introduced the subscript $l$, which means that such conical point is visited at time $t_l$. Therefore, the Riemann coordinates at the entrance of the point are written as
\begin{equation} 
v^i_{(n_l)} = c^i_{(n_l)} (t - t_l),
\end{equation}
while at the exit they are again
\begin{equation} 
\tilde{v}^i_{(n_l)} = \tilde{c}^i_{(n_l)} (t - t_l). 
\end{equation} 
Respectively, we have also for each differentiable section of the journey the relations
between the Riemann coordinate system $\{ v_{(n_l)}^i \}$ and the ordinary
coordinates $\{u^i_{(n_l)} \}$, which again can be defined as
\begin{equation}
u^i_{(n_l)} = \Phi^i_{(n_l)} (v_{(n_l)}),
\end{equation} 
and as
\begin{equation} 
\tilde{u}^i_{(n_l)} = \Phi^i_{(n_l)} (\tilde{v}_{(n_l)}) . 
\end{equation} 

At this point we have to bear in mind that in the interval $t_l < t < t_{l +1}$, the Riemann coordinate system at the exit of one point $n_l$ and that at the entrance of the next point $n_{l + 1}$ are related to each other. Our purpose now is to demonstrate the existence of a connection between two coordinates systems, one at the exit of one conical point, and the next at the entrance of the following conical point. We illustrate the geometrical significance of this procedure in Fig.~\ref{fig:4}.

Because of the equality appearing in Eq.~\eqref{eq:26}, we have that the relation between the Riemann coordinate systems is given by
\begin{equation}
v^i_{(n_{l + 1})} = c^i_{(n_{l + 1})} (t - t_{l + 1}) = \hat{R}^i_j (l + 1, l)
   \tilde{c}^j_{(n_l)} (t - t_{l + 1}),
\end{equation} 
where we have introduced a new rotation $\hat{R}^i_j (l + 1, l)$ that accounts for the fact that the Riemann coordinate system at the exit of a conical point
and the one at the entry of the next conical point are not in general aligned.
Therefore, we can write
\begin{equation}
v^i_{(n_{l + 1})} = \hat{R}^i_j (l + 1, l) \tilde{c}^j_{(n_l)} (t - t_l) - \hat{R}^i_j
   (l + 1, l) \tilde{c}^j_{(n_l)} \Delta t_l,
\end{equation} 
with $\Delta t_l = t_{l + 1} - t_l$. Finally, if we define
$\tilde{d}^j_{(n_l)} = \dot{\tilde{v}}^j_{(n_l)} \Delta t_l$, we can write the previous expression in a more compact form as
\begin{equation}
v^i_{(n_{l + 1})} = \hat{R}^i_j (l + 1, l) (\tilde{v}^j_{(n_l)} -
   \tilde{d}^j_{(n_l)}).
   \label{eq:33}
\end{equation}

And now, using the previous Eqs.~\eqref{eq:26}, we obtain
\begin{equation}
\tilde{v}_{(n_{l + 1})}^i = R^i_j (l + 1) \hat{R}^j_k (l + 1, l)
   (\tilde{v}^j_{(n_l)} - \tilde{d}^j_{(n_l)}).
\end{equation}
We note that $\tilde{d}^j_{(n_l)}$ is just the distance between the two
conical points in the Riemann coordinate chart, for which the motion is
uniform. Recalling that both matrices $R$ and $\hat{R}$ belong to the group $\tmop{SO} (2,\mathbbm{R})$, we can define a random \tmem{transition matrix} and group together both rotations in one as
\[ \tilde{R}^i_k (l + 1, l) = R^i_j (l + 1) \hat{R}^j_k (l + 1, l). \]
Now we can write the connection between two Riemann coordinates of consecutive sections as
\begin{equation}
\begin{array}{l}
     \tilde{v}_{(n_{l + 1})}^i = \tilde{R}^i_j (l + 1, l) (\tilde{v}^j_{(n_l)}
     - \tilde{d}^j_{(n_l)}).
   \end{array}
   \label{eq:35}
\end{equation}
Finally, a similar and more general relation can be found between any two conical points $k$ and $l$ along the trajectory
\begin{align}
     \tilde{v}_{(n_l)}^r &= R^r_s (l) \hat{R}^s_j (l, l - 1) \ldots . R^j_k (k + 1)
     \hat{R}^k_m (k + 1, k)  \tilde{v}_{(n_k)}^m - &&\\\nonumber &- \sum_{s = k}^{l - 1} R^r_k (l) \hat{R}_j^k (l, l - 1) \ldots R^j_k (s + 1)
     \hat{R}^k_m (s + 1, s)  \tilde{c}^m_{(n_s)} \Delta t_s .
\end{align}
Following identical arguments as before, we define the matrix $\tilde{R} (l, k)$, which relates any two conical points along the trajectory
\begin{equation}
\tilde{R}_m^r (l, k) = R^r_s (l) \hat{R}^s_j (l, l - 1) \ldots . R^j_k (k + 1) \hat{R}^k_m (k + 1, k),
\end{equation}
to obtain
\begin{equation} 
     \tilde{v}_{(n_l)}^r = \tilde{R}_m^r (l, k) \tilde{v}_{(n_k)}^m - \sum_{s = k}^{l - 1} \tilde{R}_m^r (l, s) \tilde{d}^m_{(n_s)}.
\end{equation}
Therefore, the transformation between Riemann coordinate systems of two consecutive segments of a trajectory consist of a rotation and a translation (see commutative diagram in Fig. 5), and the same holds for any other two coordinate systems. Now, after introducing the translation operator $T_d$, the relation between the ordinary coordinates $\tilde{u}_{(n_l)}^i$ at two successive conical points yields
\begin{equation}
\tilde{u}^i_{(n_{l + 1})} = \Phi_{(n_{l + 1})} \circ \tilde{R} (l + 1, l)
   \circ T_{\tilde{d}_{(n_l)}} \circ \Psi_{(n_l)}~\tilde{u}^j_{(n_l)}.
\end{equation}

\begin{figure}
\centering
  \resizebox{154pt}{124pt}{\includegraphics{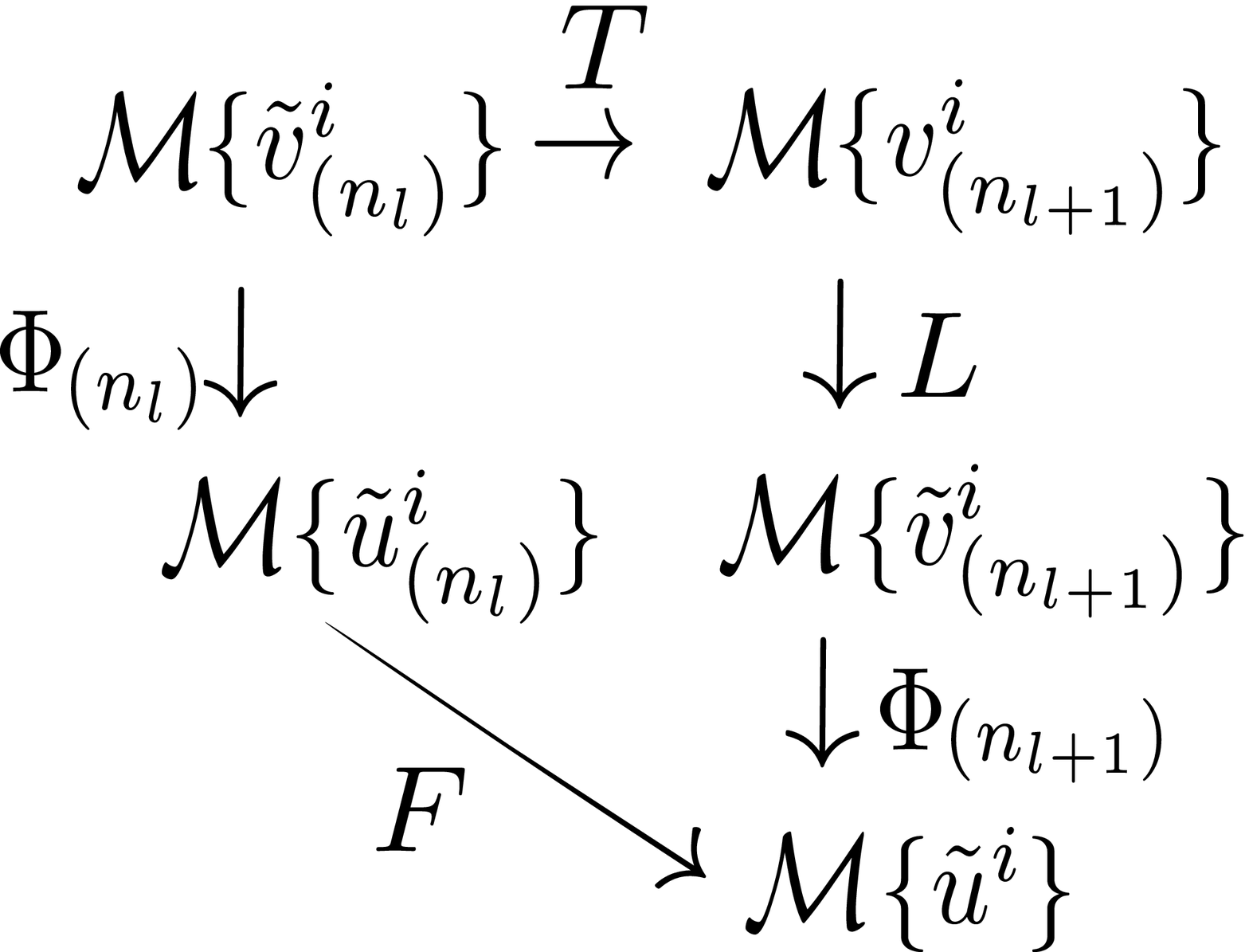}}
  \caption{\textbf{Switching coordinate systems}. A commutative diagram showing the relations among the Riemann coordinates of two successive conical points and the induced nonlinear transformation between the ordinary coordinate system on a manifold with several non-differentiable points.}
  \label{fig:5}
\end{figure}

As can be seen in the commutative diagram show in Fig.~\ref{fig:5}, the original coordinate systems are finally related by the equation
\begin{equation}
\tilde{u}^i_{(n_{l + 1})} = F^i_{(l)}(\tilde{u}_{(n_l)}),
\label{eq:40}
\end{equation}
where the random function $F_{(l)} = \Phi_{(n_{l + 1})} \circ \tilde{R}(l + 1, l)\circ T_{\tilde{d}_{(n_l)}} \circ \Psi_{(n_l)}$ has been defined. The relation appearing in Eq.~\eqref{eq:40} proves that the process is
Markovian, as it was expected, by construction. The tangent vector of the particle in the original coordinate system $\tilde{u}^i_{(n_l)}$ also obeys an iterative equation depending on one conical point and its successor. Differentiating the Eq.~\eqref{eq:35} yields
\begin{equation}
\begin{array}{l}
     \dot{\tilde{v}}_{(n_{l + 1})}^i = \tilde{R}^i_j (l + 1, l)
     \dot{\tilde{v}}^j_{(n_l)}
   \end{array}.
   \label{eq:41} 
\end{equation}
In this way, we have demonstrated that the tangent vector fluctuates stochastically as it moves through the manifold. We can write in the primary set of coordinates
\begin{equation}
\dot{\tilde{u}}^i_{(n_l)} = \dfrac{\partial \Phi^i_{(n_l)}}{\partial
   \tilde{v}^j_{(n_l)}} \dot{\tilde{v}}^j_{(n_l)} = \dfrac{\partial
   \Phi^i_{(n_l)}}{\partial \tilde{v}^j_{(n_l)}} \tilde{R}_m^j (l, l - 1)
   \dot{\tilde{v}}^m_{(n_{l - 1})},
\end{equation}
and this can be further developed as
\begin{equation}
     \dot{\tilde{u}}^i_{(n_l)} = \left( \dfrac{\partial
     \Phi^i_{(n_l)}}{\partial \tilde{v}^j_{(n_l)}} \tilde{R}_m^j (l, l - 1)
     \frac{\partial \Psi^m_{(n_{l - 1})}}{\partial \tilde{u}^k_{(n_{l -
     1})}} \right) \dot{\tilde{u}}^k_{(n_{l - 1})}.
     \label{eq:43}
\end{equation}
This relation will be of crucial importance when deriving the geodesic Langevin equations ahead. Therefore, we have proved the existence of a stochastic matrix $G$ connecting two tangent vectors belonging to two successive segments of the trajectory
\begin{equation}
G^i_k (\tilde{u}_{(n_{l - 1})}) = \dfrac{\partial \Phi^i_{(n_l)}}{\partial
   \tilde{v}^j_{(n_l)}} \tilde{R}_m^j (l, l - 1) \dfrac{\partial \Psi^m_{(n_{l-1})}}{\partial \tilde{u}^k_{(n_{l - 1})}}.
\end{equation}
This mapping can also be written as
\begin{equation}
G^i_k (\tilde{u}_{(n_{l - 1})}) = \dfrac{\partial F^i_{(l)}}{\partial
\tilde{u}^k_{(n_{l - 1})}},
\end{equation}
what demonstrates that the tangent vector also obeys a Markov process, in the form
\begin{equation}
\dot{\tilde{u}}^i_{(n_l)} = G^i_k (\tilde{u}_{(n_{l - 1})})
\dot{\tilde{u}}^k_{(n_{l - 1})}.
\end{equation}
To delve deeper into the understanding of the nature of the random walk, we recast the probability distributions. The stochastic coordinate system can be represented by a probability distribution and a probability density function of the coordinates can be determined.

If we assign a probability density $p_V (\tilde{v}_{(n_l)}, t_l)$ to a
certain conical point, we can write a \emph{recurrence equation} for the probability
function at different time instants between two conical points as
\begin{equation}
p_V (\tilde{v}_{(n_{l + 1})}, t_{l + 1}) = \sum_{n_l = 1}^N p (\tilde{R} (l
   + 1, l)) p_V (\tilde{v}_{(n_l)}, t_l),
\end{equation}
where the mutual independence of the stochastic coordinates has been assumed.
Note that the coordinates $\tilde{v}_{(n_{l + 1})}$ at time $t_{l + 1}$ can
always be connected to the previous conical point $\tilde{v}_{(n_l)}$ by means of a rotation and a translation (see again Eq.~\eqref{eq:35}), and that coordinates are random variables in this fractal mechanics. Since the determinant of the Jacobian of these two transformations is equal to one, the relation can be finally written as
\begin{equation}
p_V (\tilde{v}_{(n_{l + 1})}, t_{l + 1}) = \sum_{n_{l} = 1}^N p
   (\tilde{R} (l + 1, l)) p_V (\tilde{v}_{(n_{l+1})} + \tilde{d}_{(n_{l})}, t_l).
   \label{eq:48}
\end{equation}
We briefly consider an example to clarify this point. Assume that an infinite number of conical points are distributed in $\mathbbm{R}^2$ on a cubic lattice and that, for simplicity, only first-neighbor geodesics are allowed. In this case, there are only four possible matrices to be chosen at a conical point $\tilde{R} (l + 1, l) \in \{ R (0), R (\pi / 2), R (\pi), R (3 \pi / 2) \}$, which we recall that are chosen at random. The four rotations constitute the set of possible events at the conical point. Assuming that space is \tmem{homogeneous} and \tmem{isotropic}, the random walk can not be biased, and we are led to the equiprobability assumption $p (\tilde{R} (l + 1, l)) = 1 / 4$. In this manner, we obtain the finite difference operator
\begin{equation}
\begin{array}{l}
     p_V (v_{(n_{l + 1})}, t_{l + 1}) =\\
     = \frac{1}{4} p_V (v_{(n_{l + 1})} - \tilde{d}^1_{(n_{l})}, t_l) +
     \frac{1}{4} p_V (v_{(n_{l + 1})} + \tilde{d}^1_{(n_{l })}, t_l) \\
     + \frac{1}{4} p_V (v_{(n_{l + 1})} - \tilde{d}^2_{(n_{l})}, t_l) +
     \frac{1}{4} p_V (v_{(n_{l + 1})} + \tilde{d}^2_{(n_{l})}, t_l),
   \end{array}
   \label{eq:49}
\end{equation}
that frequently appears in the treatment of diffusion processes. This recurrence equation allows to construct the master equation of the Markovian stochastic process \cite{scott}. As we show in the following section, taking a limit to the continuum of Eq.~\eqref{eq:49} also gives rise to a partial differential equation for the probability density, the so-called Fokker-Planck equation. Note that the distribution previously mentioned $p (\tilde{R} (l + 1, l))$ is just the \tmem{conditional probability density} (the propagator, in physical parlance) of the process. Mathematically, we can write
\begin{equation}
p_V (\tilde{v}_{(n_{l + 1})}, t_{l + 1} | \nobracket \tilde{v}_{(n_l)},t_l) = p (\tilde{R} (l + 1, l)).
\end{equation}

To conclude this part of the work, we prove that another process occurs for the original coordinate system, by simply performing a change of variables in the probability distribution. On the left-hand side of Eq.~\eqref{eq:48} we get
\begin{equation}
p_V (\tilde{v}_{(n_{l + 1})}, t_{l + 1}) = p_U (\tilde{u}_{(n_{l + 1})},
   t_{l + 1}) \det \tilde{J}_{(n_{l + 1})},
\end{equation}
where the determinant of the Jacobian is again given by
\begin{equation}
\det \tilde{J}_{(n_{l + 1})} = \left| \frac{\partial (\Phi^1_{(n_{l + 1})},
   \Phi^2_{(n_{l + 1})})}{\partial (\tilde{v}^1_{(n_{l + 1})},
   \tilde{v}^2_{(n_{l + 1})})} \right|_{\tilde{v}_{(n_{l + 1})}}.
\end{equation}
Note that the probability density of the Riemann coordinate system has been denoted by means of the subindex $V$, whereas in the ordinary coordinates it is now indicated by using the subindex $U$. On the right-hand side of Eq.~\eqref{eq:48} the problem is more difficult. We have that
\begin{equation}
\begin{array}{l}
     p_V (\tilde{v}_{(n_{l})}, t_{l + 1}) = p_U
     (\tilde{u}_{(n_{l})}, t_{l + 1}) \det
     \tilde{J}_{(n_{l})}
   \end{array}
\end{equation}

Thus, in the original coordinate system we have a different relation, and it
is given by
\begin{equation}
p_U (\tilde{u}_{(n_{l + 1})}, t_{l + 1}) = \sum_{n_l = 1}^N p_U (\tilde{R}
   (l + 1, l)) \frac{\det J_{(n_{l})}}{\det J_{(n_{l + 1})}} p_U
   (\tilde{u}_{(n_{l + 1})} +\tilde{\delta}_{(n_{l})}, t_l),
\end{equation}
where we have defined
\begin{equation}
\tilde{\delta}_{(n_{l})} = F_{(l)}^{-1}(\tilde{u}_{(n_{l+1})})
- \tilde{u}_{(n_{l + 1})}.
\end{equation}
As it is evident, this implies a difference in the Fokker-Planck equations related to the two fundamental coordinates systems, as will be show in the examples.

In short, we have demonstrated that the dynamics of a particle in a space with many non-differentiable points can be represented by a sequence of coordinate systems, which are related among them through random functions. Such sequence of randomly interconnected coordinate systems defines a \tmem{random walk}, and we can therefore adopt the perspective in which the whole process is represented by a single \emph{stochastic coordinate system}. As we shall see very soon, this change in the point of view of the dynamics will play a key role in the formulation of the geodesic stochastic differential equations that govern the dynamics of a moving body in a fractal space. Such equations are derived and thoroughly described right ahead.

\subsection{An everywhere non-differentiable manifold}

Things complicate noticeably if the space is everywhere non-differentiable.
Although known for a long time before the work of Mandelbrot \cite{mand}, such
geometrical objects were regarded with skepticism at first. As it is well known, they are familiarly known as \tmem{fractals} nowadays. Thanks to his work, such examples have plagued the scientific literature and revealed very useful, if not indispensable, to provide an accurate description of many natural objects, including self-similar structures for which the hypothesis of smoothness seems too restrictive or unpractical. Fractals are also essential to the description of the chaotic dynamics of many physical systems \cite{macd,sanj} and, since chaos is a universal phenomenon \cite{mott}, so are fractal structures \cite{mott2}.

As we are about to demonstrate, a possible characterization of the dynamics of a particle in a fractal space can be given by finding a Langevin equation of motion on an average manifold, where this last differentiable structure results from smoothing out the fluctuations of the rough space. These equations are differential equations in the Ito's sense and give rise to continuous trajectories with some particular Hausdorff dimension \cite{falc}, which is generally greater than their topological dimension. Once the stochastic differential equations of motion are derived, a Fokker-Planck equation describing the dynamics of the probability distribution can be worked out, if desired. Therefore, in the following lines, we concentrate our efforts on the study of the resulting continuous stochastic process and the derivation of the Langevin equations, to characterize the motion of a particle in an everywhere non-differentiable manifold.

However, before developing these concepts, one remark is unavoidable, since a fractal space presents a particular difficulty. Namely, the distance between any two points is infinite in the Lebesgue sense and, therefore, a point particle in a fractal space should not move at all in a finite time interval. However, this problem disappears if we consider that particles are physical ($i.e.$ the particles have a finite size), rather than mathematical point masses \cite{lope,lop202}. Then, we can assume that the non-differentiability of the space is not felt by the particle at scales much smaller than its size. Consequently, we can make use of the concepts exposed in the previous sections, as long as we can work a differentiable structure out of the everywhere rough original space. The question is thus how this averaging process has to be carried out, to construct a smoothed space. Certainly, there exist many smoothing techniques in the literature but, perhaps, the methodology that suits more naturally our problem is a method known as kernel smoothing \cite{wand}. This method convolves the fractal or fluctuating function to be smoothed with a kernel function $K_{\epsilon}$ of typical size $\epsilon$. For the sake of clarity, we can consider as an example a Gaussian kernel
\begin{equation}
K_{\epsilon} (u, v) = \frac{1}{\sqrt{4 \pi \epsilon^2}} e^{- \frac{u^2 +
   v^2}{4 \epsilon^2}}.
\end{equation}
   
Since the size of the particle marks the limiting value below which the system can be considered smooth, such particle's size establishes approximately the $\epsilon$ value of the kernel. In other words, given some parametrization of the fractal manifold $u^i$ and a position vector on it $\tmmathbf{x}$, the smoothed position vector $\tmmathbf{x}_{\epsilon}$ is defined in the smoothed manifold by means of the convolution
\begin{equation}
\tmmathbf{x}_{\epsilon} (u_1, u_2) = \int^{\infty}_{- \infty}
   \int^{\infty}_{- \infty} K_{\epsilon} (u - u_1, v - u_2) \tmmathbf{x} (u,
   v) d u d v.
\end{equation}

Of course, for an originally smooth manifold we can always consider the limit $\epsilon \rightarrow 0$ and obtain the equivalence $\tmmathbf{x}_{\epsilon} (u_1, u_2) \rightarrow \tmmathbf{x} (u_1, u_2)$, since $K_{\epsilon}(u,v) \rightarrow \delta(u,v)$. It is precisely in the average manifold that all the forthcoming tensors and symbols have to be computed. In other words, we can think of the dynamics on a fractal space as a stochastic movement in a base smooth manifold, and the fractal space itself can be represented as a smooth manifold on which fluctuations are incorporated.

Just as an enlightening metaphor, we can think of a solid particle moving in a fractal space as a ship traveling over the seas. If the particle is big
enough, then we can think of a huge vessel on a calmed day, in which, having
fixed the direction of the rudder and the velocity of the vessel, it moves uniformly, because the fluctuations of the water are not strong enough to alter its trajectory. On the other hand, for a small particle that feels the fractality of the space, we should rather think of a small vessel on a stormy day, in which the very complex waves and fluctuations of the space deviate it ceaselessly from its direction.

We now proceed to characterize the process continuously, using Ito's
stochastic calculus. As we have previously said, our main purpose is to use
the theory of continuous stochastic processes to derive a system of \emph{Langevin geodesic equations}. This goal is achieved by considering a continuous limit of the equations obtained in the previous section. Starting from the discrete
equation
\begin{equation}
\tilde{v}_{(n_{l + 1})}^i = \tilde{R}^i_j (l + 1, l) (\tilde{v}^j_{(n_l)} -
   \tilde{c}^j_{(n_l)} \Delta t_l),
\end{equation}
we obtain its continuous associated process by introducing infinitesimal
variations. This allows to relate two Riemann coordinate systems as
\begin{equation}
\tilde{v}^i (t + d t) = \tilde{R}^i_j (t + d t, t) (\tilde{v}^j_{} (t) -
   \tilde{c}^j (t) d t),
\end{equation}
which can be written in the form
\begin{equation}
\tilde{v}^i (t + d t) = \tilde{R}^i_j (t + d t, t) (\tilde{v}^j_{} (t) -
   \dot{\tilde{v}}^j (t) d t).
\end{equation}
The infinitesimal rotation can be represented by the angle $\alpha
(t)$, which is a random function. For simplicity, we assume that this function obeys a continuous Wiener process $W(t)$. This process is a Gaussian random variable evolving in time with expected value $\langle W(t) \rangle=0$ and independent increments, which means that the condition $\langle (W(t_{4})-W(t_{3}))((W(t_{2})-W(t_{1})) \rangle=0$ holds whenever $[t_{1},t_{2}] \cap [t_{3},t_{4}]= \emptyset$. Neglecting terms of order greater that $d t$, we obtain
\begin{equation}
\tilde{R}^i_j (t + d t, t) \equiv \tilde{R}_j^i (d \alpha (t)) = \delta^i_j
   \left( 1 - \frac{(d \alpha (t))^2}{2} \right) + d \alpha (t)
   \varepsilon^i_j,
   \label{eq:62}
\end{equation}
where $\varepsilon^i_j$ is the infinitesimal generator of rotations in the
plane, which we refer as the {\tmem{spin}} hereafter. Substitution yields
\begin{equation}
\dot{\tilde{v}}^i_{} (t) = - \frac{1}{2} \frac{(d \alpha)^2}{2 d t}
   \tilde{v}^i_{} (t) + \frac{d \alpha}{d t} \varepsilon^i_j \tilde{v}^j_{}
   (t) .
\end{equation}
   
We are considering a Wiener process in the form $d \alpha = \sqrt{2 D} d W$, where the differential Wiener increments $d W (t)=W(t+d t)-W(t)$ have been defined. These increments are sometimes written in terms of the normal distribution $N(\mu,\sigma^2)$, where $\mu$ is the expected value and $\sigma^2$ the variance \cite{lemo}. Using this notation, we have $d W = N_t^{t + d t} (0, d t) = \sqrt{d t} N_t^{t + dt} (0, 1)$, with $N_t^{t + d t} (0, 1)$ the normal distribution with mean value zero and variance equal to one in the interval $[t,t+dt]$. The parameter $D$ measures the intensity of the fluctuations or, in the present analysis, how much the angle of the tangent vector fluctuates. Then, it can be demonstrated that $(d W)^2 = d t$, a result well known as Ito's rule  \cite{jaco}.  It is for this reason that we have to keep terms up to second order in $d \alpha$ when expanding in series the Eq.~\eqref{eq:62}. Therefore, we have the stochastic differential equation
\begin{equation}
\dot{\tilde{v}}^i_{} (t) = - D \tilde{v}^i_{} (t) + \sqrt{2 D}
   \varepsilon^i_j \tilde{v}^j_{} (t)  \dot{W}.
\end{equation}

The derivative of the Wiener process is precisely a white noise $\eta (t)$
and, defining the angular velocity as $\omega (t) = \sqrt{2 D} \eta (t)$, the whole process can be written as
\begin{equation}
\dot{\tilde{v}}^i_{} (t) = - D \tilde{v}^i_{} (t) + \omega (t)
   \varepsilon^i_j \tilde{v}^j_{} (t).
\end{equation}

The inconvenience of this stochastic differential equation is that the
resulting process has an infinite kinetic energy in the limit $d t \rightarrow
0$, since $\omega (t) \sim \dot{W} \sim 1 / \sqrt{d t}$. In mathematical
terms, and assuming that the mass of the particle is equal to unity, this is
written as
\begin{equation}
K = \frac{1}{2}  \dot{\tilde{v}}_i \dot{\tilde{v}}^i_{} = \frac{1}{2} (D^2
   + \omega^2 (t)) \tilde{v}_i \tilde{v}^i_{}.
\end{equation}
   
This hurdle disappears if we use the Eq.~\eqref{eq:41} for the tangent vectors
directly. In this case we start from the more simple equation
\begin{equation}
\dot{\tilde{v}}_{(n_{l + 1})}^i = \tilde{R}^i_j (l + 1, l)\dot{\tilde{v}}^j_{(n_l)},
\end{equation}
and taking the continuous limit we derive
\begin{equation} 
\dot{\tilde{v}}^i (t + d t) = \tilde{R}^i_j (d \alpha) \dot{\tilde{v}}^j(t),
\end{equation}
what yields the stochastic differential equation
\begin{equation} 
d \dot{\tilde{v}}^i (t) = - D \dot{\tilde{v}}^i (t) d t + \sqrt{2 D}
   \varepsilon^i_j \dot{\tilde{v}}^j_{} (t) d W.
\end{equation}

The result is basically a process of \tmem{elastic scattering} which, as it is well-known, \tmem{conserves energy}. The previous equation just says that the non-differentiability of the space acts on the particle by dispersing it. If we recuperate the scenario of the previous section, we can consider that the non-differentiable points are equivalent to scattering centers which act on the particle by rotating its tangent vector. Concerning the original coordinates of the manifold, as appearing in Eq.~\eqref{eq:43}, we have the continuous equation
\begin{equation} 
\dot{\tilde{u}}^i (t + d t) = \left( \dfrac{\partial \Phi^i}{\partial
   \tilde{v}^j} (t + d t) \tilde{R}_m^j (d \alpha (t)) \dfrac{\partial
   \Psi^m}{\partial \tilde{u}^k} (t) \right) \dot{\tilde{u}}^k (t) . 
\end{equation} 
This equation, when expanded in series on its left-hand side yields
\begin{equation}
\dot{\tilde{u}}^i (t) + \ddot{\tilde{u}}^i (t) d t = \left( \frac{\partial
   \Phi^i}{\partial \tilde{v}^j} (t + d t) \tilde{R}_m^j (d \alpha (t))
   \frac{\partial \Psi_m}{\partial \tilde{u}^k} (t) \right)
   \dot{\tilde{u}}^k (t).
\end{equation}
If we now expand the rotation as done in the previous lines, we obtain
\begin{equation}
    \dot{\tilde{u}}^i (t) + \ddot{\tilde{u}}^i (t) d t
    = \frac{\partial \Phi^i}{\partial \tilde{v}^j} (t) \tilde{R}_m^j (d\alpha (t)) \frac{\partial \Psi^m}{\partial \tilde{u}^k} (t)
    \dot{\tilde{u}}^k (t) + \frac{\partial^2 \Phi^i}{\partial \tilde{v}^j \partial \tilde{v}^l} \dot{\tilde{v}}^l (t) d t \tilde{R}_m^j (d \alpha (t)) \frac{\partial \Psi^m}{\partial \tilde{u}^k} (t) \dot{\tilde{u}}^k (t),
\end{equation}
which gives as a result
\begin{flalign}
    \dot{\tilde{u}}^i (t) + \ddot{\tilde{u}}^i (t) d t &= \frac{\partial \Phi^i}{\partial \tilde{v}^j} (t) \frac{\partial
    \Psi^m}{\partial \tilde{u}^k} (t) \delta^j_m \left( 1 - \frac{(d
    \alpha)^2}{2} \right) \dot{\tilde{u}}^k(t) + &&\\\nonumber 
    &+ d \alpha (t) \frac{\partial \Phi^i}{\partial \tilde{v}^j} (t) \varepsilon_m^j \frac{\partial \Psi^m}{\partial \tilde{u}^k}(t) \dot{\tilde{u}}^k (t)  + \frac{\partial^2 \Phi^i}{\partial \tilde{v}^j \partial \tilde{v}^l} \frac{\partial
    \Psi^l}{\partial \tilde{u}^s}(t) \delta^j_m \frac{\partial
    \Psi^m}{\partial \tilde{u}^k}(t) \dot{\tilde{u}}^s (t)
    \dot{\tilde{u}}^k (t) d t.
\end{flalign}
In Appendix C we prove that the last term can be related to the Christoffel symbols of the average manifold, while the penultimate term can be computed from the first fundamental form $g$. The result is
\begin{equation}
\ddot{\tilde{u}}^i (t) d t = - \frac{d \alpha^2}{2} \dot{\tilde{u}}^k (t) +
   \Omega^i_k (\tilde{u}) \dot{\tilde{u}}^k (t) d \alpha (t) - \Gamma^i_{s k}(\tilde{u}) \dot{\tilde{u}}^s (t) \dot{\tilde{u}}^k (t) d t.
\end{equation}
The matrix $\Omega^i_k$ is related to the spin on a curved surface and is written as
\begin{equation}
\Omega = \frac{1}{\sqrt{|g|}} H, H = \varepsilon g = \left( \begin{array}{ll}
     -g_{12} & - g_{22}\\
     g_{11} & g_{21}
   \end{array} \right),
\end{equation}
with $g$ is the metric tensor. If the rotation is a Wiener process, we finally obtain the stochastic differential equation in the Ito's sense
\begin{equation}
\begin{array}{l}
     d \dot{\tilde{u}}^i (t) + \Gamma^i_{s k} (\tilde{u}) \dot{\tilde{u}}^s
     (t) \dot{\tilde{u}}^k (t) d t = - D \dot{\tilde{u}}^i (t) d t +
     \Omega^i_k (\tilde{u}) \dot{\tilde{u}}^k (t) \sqrt{2 D} d W,
   \end{array}
\end{equation}
which can be more clearly written by giving a phase space representation and
using conventional letters for the coordinates. Thus, we now name the space
coordinates $x^i$, as it is customary done in the study of classical physics,
and use the letter $v^i$ to represent the velocities. This change in the
notation leads to the stochastic differential equation
\begin{equation}
  \begin{array}{l}
    d x^i = v^i d t\\
    d v^i = - D v^i d t + \Omega^i_k (x) v^k \sqrt{2 D} d W - \Gamma^i_{s
    k} (x) v^s v^k d t,
  \end{array}
\end{equation}
which constitutes the main result of the present work. Note that in the limit $D \rightarrow 0$ we recover the geodesic equations corresponding to a smooth manifold. Therefore, the theory of smooth manifolds is contained in the present fractal mechanics. We recall that the tensor $\Omega^i_k(x)$ and the symbols $\Gamma^i_{s k} (x)$ have to be computed on the {\tmem{averaged manifold}} and that, although it has been omitted for convenience in the notation, the coordinates themselves depend on the scale $\epsilon$. This fact is not only a mere question of particle size, but it is of the greatest importance concerning the nature of a particle. For if we assume that the extended particle is allowed to change its shape, each of its points tending to follow a geodesic fractal path, the geometry of the particle is permanently evolving. This is well known even for smooth manifolds, where the Riemann curvature tensor measures the change of size and shape of such a body. This dependence of the physical laws on the scale poses numerous questions concerning the relativity of scales and the dynamics among the different scales. These fundamental issues have been very brilliantly discussed somewhere else but in a different mathematical context \cite{nott1}.

\section{Two illustrative examples}

We now study numerically two particularly simple situations, that will serve to illustrate
the application of this theory of fractal motion in a clear and straightforward fashion. As a first example, we shall consider a space that is flat on the average, while then a smoothed space with positive constant curvature will be described.

\subsection{A fractal space that is flat on average}

As a first example, we consider a space that, when averaged at the scale of
interest, is completely flat. In this case the Christoffel symbols of the
average manifold are zero $\Gamma_{i j}^k = 0$. On the other hand, since the
metric tensor is euclidean, we have that $\Omega$ is the infinitesimal
generator of two-dimensional rotations. Therefore, we obtain the set of differential
equations
\begin{equation}
\begin{array}{l}
     d x = v_x d t\\
     d v_x^{} = - D v_x d t - v_y \sqrt{2 D} d W\\
     d y = v_y d t\\
     d v_y = - D v_y d t + v_x \sqrt{2 D} d W
   \end{array} .
\end{equation}

\begin{figure}
\centering
  \resizebox{250pt}{200pt}{\includegraphics{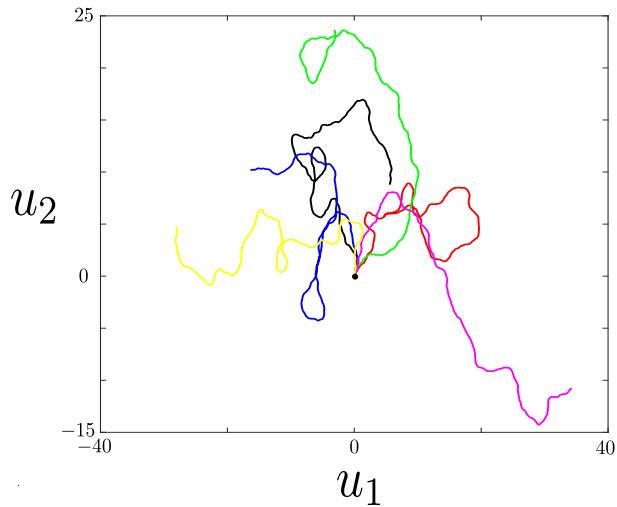}}
  \caption{{\tmstrong{Random Walks.}} A set of projected geodesic paths on a fractal surface whose average at the relevant scale is a plane. A value of $D=0.1$ has been considered for these simulations. The random nature of these various walks can be clearly appreciated. A Milstein scheme has been used to integrate the stochastic differential equations, where the step of integration has been set to $h=0.01$.}
  \label{fig:6}
\end{figure}

\begin{figure}
\centering
  \resizebox{400pt}{200pt}{\includegraphics{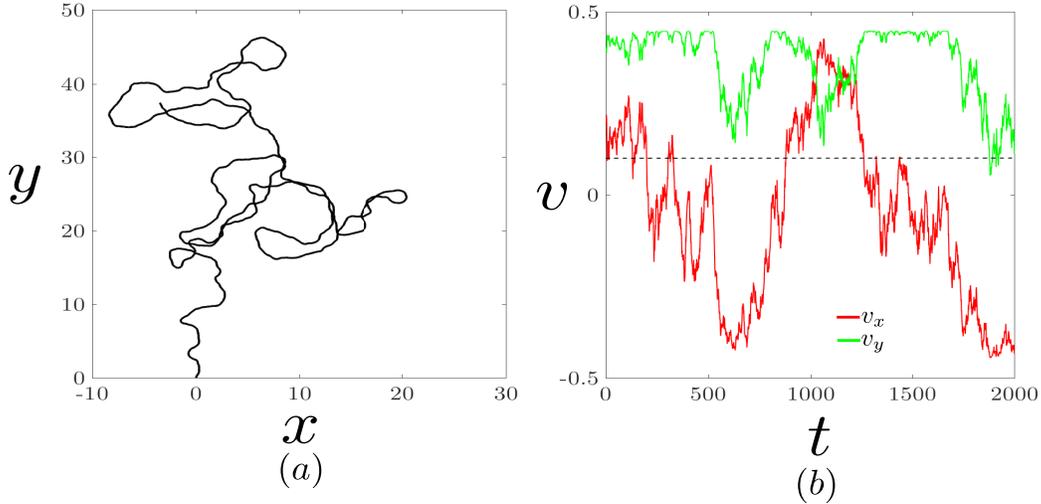}}
  \caption{{\tmstrong{Fractal trajectories.}} (a) A random walk on a space that is flat on average, with fluctuations of size $D=0.1$. The initial conditions are $(x,y)=(0,0)$, while the initial momentum is $(v_x,v_y)=(0.2,0.4)$. A Milstein scheme has been used to integrate the stochastic differential equations, where the step of integration has been set to $h=0.01$. (b) The time series of the two components of the momentum of the particle, showing that fluctuations arise as a consequence of the fractal nature of space. The resulting curve exhibits the usual self-similar structure of noisy series, as can be verified by zooming in a small time interval. The dotted line represents the kinetic energy of the particle, which is clearly conserved.}
    \label{fig:7}
\end{figure}

These equations have appeared in other contexts of physics, as thoroughly
discussed in Sec.~VI. In any case, they represent a process of
elastic scattering, so that the particle is permanently changing the direction
of the velocity vector without changing its modulus. We show some simulations
of the trajectory of the particle in Fig.~\ref{fig:6}. As can be seen, the effect of the non-smoothness of the space is to deviate the particle from its
natural differentiable geodesic movements, producing the diffusion of the
moving corpuscle. The time series of the momentum for a typical journey in this average flat space are shown in Fig.~\ref{fig:7}. As can be seen, the momentum performs considerable fluctuations at all scales, manifesting the self-similar nature of the underlying Wiener process.

A Fokker-Planck equation can be derived from the previous system of ordinary
differential equations for a probability distribution defined on the phase
space $P (x, y, v_x, v_y, t)$. However, since the previous stochastic
differential equations are only partially coupled, we can consider more simply
a partial differential equation on $P (v_x, v_y, t)$. This partial
differential equation is of the form
\begin{equation} \frac{\partial P}{\partial t} = D \frac{\partial}{\partial v_i} (v_i P) + D
   \left( v_y \frac{\partial}{\partial v_x} - v_x \frac{\partial}{\partial
   v_y} \right)^2 P.
\end{equation}
Therefore, we see the tendency of the system to rotationally diffuse the
probability distribution in the velocity space, together with the limitation
in the velocity fluctuations imposed by the first terms. This picture is even
clearer if we consider polar coordinates $(v \cos \theta, v \sin \theta)$ in
the velocity space, instead of the Cartesian pair $(v_x, v_y)$. With this
change of coordinates we can transform to the much more simple partial differential equation
\begin{equation} \frac{\partial P}{\partial t} = D P + D \frac{\partial}{\partial v} (v P) +
   D \frac{\partial^2 P}{\partial \theta^2}.
\end{equation}

Except for the first term, this equation is very similar to the
Ornstein-Uhlenbeck process, and it can be easily solved using Fourier's transform and the method of characteristics. Note that the diffusion occurs in the angular coordinate (rotational diffusion) and that a particle initially at rest remains at rest in its position in the present model.

\subsection{A fractal space that has constant curvature on average}

\begin{figure}
\centering
  \resizebox{400pt}{200pt}{\includegraphics{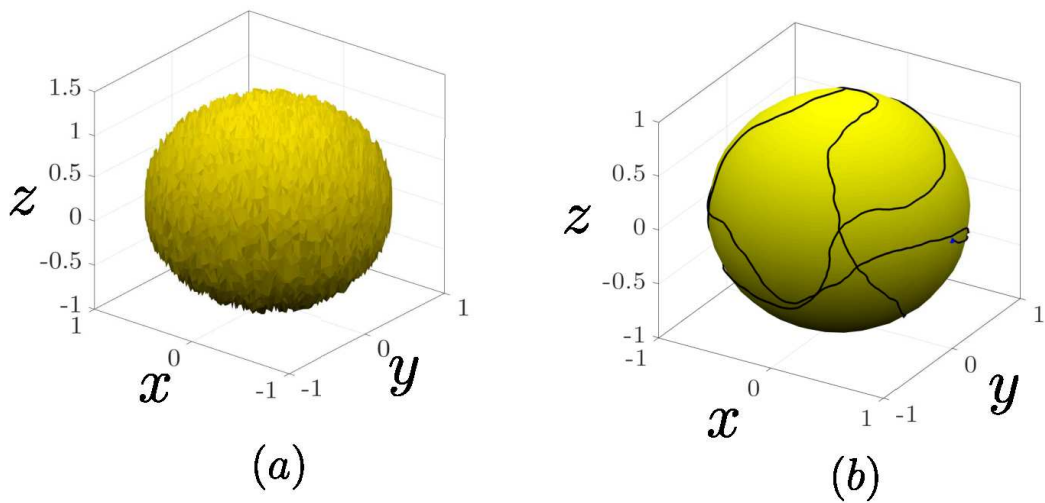}}
  \caption{{\tmstrong{Fractal curved space.}} (a) A rough sphere displaying a very complicated shape. (b) A trajectory on the smooth manifold $S^2$, where the value of $D=0.5$ has been used. A Milstein scheme has been used to integrate the stochastic differential equations, where the step of integration has been set to $h=0.01$.}
    \label{fig:8}
\end{figure}

As a second example, we study a smoothed space that is not flat, but
presents positive constant curvature. We consider the sphere as the averaged ideal manifold (see Fig.~\ref{fig:8}), where the conventional spherical charts $(\theta,\phi)$ are given. We compute in the first place the matrix $\Omega$. We
have that
\begin{equation}
H = \left( \begin{array}{ll}
     0 & - \sin^2\theta\\
     1 & 0
   \end{array} \right)
\end{equation}
which yields the matrix
\begin{equation}
\Omega = \left( \begin{array}{ll}
     0 & - \sin \theta\\
     \csc \theta & 0
   \end{array} \right).
\end{equation}
This allows to write the differential equations
\begin{equation}
\begin{array}{l}
     d \theta = \omega_{\theta} d t\\
     d \omega_{\theta} = - D \omega_{\theta} d t - \sin \theta \omega_{\phi}
     \sqrt{2 D} d W + \sin \theta \cos \theta \omega_{\phi} \omega_{\phi} d
     t\\
     d \phi = \omega_{\phi} d t\\
     d \omega_{\phi} = - D \omega_{\phi} d t + \csc \theta \omega_{\theta}
     \sqrt{2 D} d W - 2 \cot \theta \omega_{\phi} \omega_{\theta} d t
   \end{array}.
\end{equation}

\begin{figure}
\centering
  \resizebox{400pt}{200pt}{\includegraphics{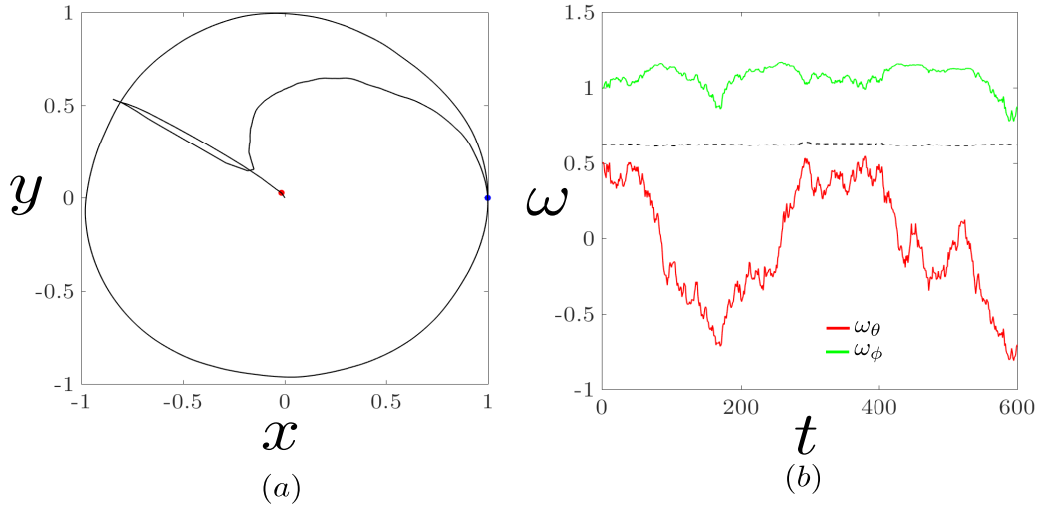}}
  \caption{{\tmstrong{Fractal trajectories.}} (a) A random walk on a space that has a constant curvature scalar ($R=2$) on average with $D=0.05$. The trajectory has been projected onto the $xy$-plane for clarity. The particle is initially set on the equator $(\theta,\phi)=(\pi/2,0)$  (blue dot) with angular momentum $(\omega_\theta,\omega_\phi)=(0.5,1.0)$. A Milstein scheme has again been used with $h=0.01$. The algorithm stops when one of the poles is reached (red dot), were a switch of charts is required. (b) The time series of the two components of the angular velocity of the particle, showing the fluctuations that arise as a consequence of the fractal nature of the space. Again, the dotted line represents the kinetic energy of the particle, which is conserved.}
    \label{fig:9}
\end{figure}

It is interesting to note the effect of nonlinearity in this case. We compute different trajectories in Fig.~\ref{fig:8}, for different values of $D$. As the particle travels along some geodesic, its tangent vector fluctuates and switches to some other geodesic. This occurs all the time, and since the average manifold presents curvature, a feedback phenomenon is permanently operating between nonlinearity and fluctuations, the former tending to amplify or dissipate the latter. When the geodesics curvature is positive and parallel geodesics tend to get closer, fluctuations are reduced. On the other hand, for negative curvature spaces parallel geodesics tend to deviate, and the opposite situation is found, giving rise to the amplification of fluctuations. In Fig.~\ref{fig:9} the trajectory of a particle launched with some inclination towards the south pole is seen. The particle performs a whole trip around the sphere and then the fluctuations in the angular momentum of the particle drive toward the south pole. These fluctuations are represented in Fig.~\ref{fig:9}(b), which clearly evince the self-similar nature of the Wiener process.

\section{Discussion}

We have developed a dynamical theory that allows representing the motion of a particle in complicated physical situations, where the medium can have average intrinsic curvature and exhibits unpredictable fluctuations. These fluctuations can be incorporated in our mathematical conception of space, its main effect being the systematic scattering of the particle, which is permanently deviated from its differentiable geodesics. Moreover, given an initial condition, we see that a fractal space introduces an infinite set of geodesics, as has been claimed in previous works \cite{nott1}.

In our general treatment, the motion of the body is described using Ito's
calculus, employing a stochastic differential equation in an averaged phase space. This contrasts with other theories, which are formulated in configuration space \cite{nels2} and, in this respect, is in closer resemblance to other previous approaches to rough spaces \cite{kuri}. This occurs because we have introduced the fluctuations in the tangent vector using Riemann coordinates, what leads to a covariant formulation of the theory. From a physical point of view, this is certainly much more natural if there is an inherent curvature in the averaged space. Nevertheless, our formalism is very robust and permits the usage of other symmetry groups in replacement of the orthogonal group, which can account for dissipative or even excitable processes \cite{lope,lop202}. Therefore, inelastic scattering processes can be also modeled in the present formalism. If, in addition to rotations, the translation of tangent vectors is allowed, a reduction to configuration space by considering a Smoluchowski's limit can be attained, which leads to a stochastic differential in configuration space, similar to those appearing in Nelson's work \cite{nels2}

Higher dimensional spaces are easy to implement as well (\emph{e. g.} Lorentzian
manifolds), since our description has been carried out bearing in mind a general scheme. Again, the only requirement is to use appropriate representations of symmetry groups for the stochastic dynamical bundle. We also note that, since our formulation is made in terms of classical physical concepts, external fields can be easily incorporated as well in the stochastic geodesic equations. Even the assumed Markovian stochastic processes can be extended to processes with memory using other types of Brownian motion, for example a fractional Brownian motion \cite{mand2}. Certainly, the Markovian choice has a fundamental advantage, since a Langevin equation can be easily translated into a Fokker-Planck description, from which a conditional probability can be sometimes computed. Once the propagator has been worked out, it is immediate to advance any initial distribution by convolution.

We close this work discussing a few possible physical applications of the
present formalism. In the first place, it is evident that elastic scattering
processes are easy to implement. For example, electrons colliding with Helium ions in a hot plasma, or neutrons diffusing through a matrix of cold material are two possible candidates \cite{lemo}. Interestingly, scattering processes in non trivial manifolds can be afforded now. For instance, electron scattering in conductors with arbitrary complicated geometries. Processes of diffusion on fractals \cite{osha} or diffusion in heterogeneous media, can be adequately modeled as well. Finally, examples from the biological realm can be drawn too, as the migratory motion of animal species \cite{bril}, which follow geodesic paths but deviate briefly from them when foraging. The motion of flagellated bacteria has also been pointed out \cite{berg}.

In summary, we believe that this new formalism might serve to extend the phase
space description of differentiable manifolds to more complicated motions commonly found in nature, where unexpected changes appear in the trajectories of a natural body. In these contexts, the traditional approach based on smooth manifolds would be unpractical, because the fast fluctuations do not permit an easy representation of the tremendously chaotic differentiable trajectories. In this way, a road towards a dynamical description of fractal trajectories in nature is paved. Although implicitly present in the theory of stochastic processes, which is here used instrumentally, this formulation of fractal motion is, from the point of view of differentiable geometry, more general and useful, since it is a covariant formulation. It was perhaps lacking in the literature in comparison to the approaches previously indicated. We hope that it might serve to enlarge the wide application of classical physical concepts to fractal geometries in the description of natural evolving processes in space and time.

\section{ACKNOWLEDGMENTS}
The author wishes to thank Sijo K. Joseph for fruitful discussion on the concept of a fractal space and for his advice on the elaboration of the present manuscript.

\begin{appendices}
\renewcommand{\theequation}{A\arabic{equation}}
\setcounter{equation}{0} 

\section{Riemann normal coordinates on a cone}

Our aim in this section is to obtain the Riemann coordinates of a cone with
apex at $(0, 0, 0)$ and generatrix $(0, 1, a)$, at the point $\left( 1 /
\sqrt{1 + a^2}, 0, a / \sqrt{1 + a^2} \right)$. The position vector can be
written as
\begin{equation}
\tmmathbf{x} (u, v) = (u \cos v, u \sin v \nocomma, a u).
\end{equation}

The tangent basis is computed as
\begin{equation} 
\tmmathbf{x}_u = \frac{\partial \tmmathbf{x}}{\partial u} = (\cos v, \sin v
   \nocomma, a), \tmmathbf{x}_v = \frac{\partial \tmmathbf{x}}{\partial v} =
   (- u \sin v, u \cos v \nocomma, 0),
\end{equation}
and from these we can compute the first fundamental form in a simple manner
$g_{i j} =\tmmathbf{x}_{u_i} \cdot \tmmathbf{x}_{u_j}$. The result is
\begin{equation}
g = (1 + a^2) d u^2 + u^2 d v^2.
\end{equation}

We now rescale the coordinates to obtain the typical metric for a polar coordinate system
\begin{equation}
u_1 = \sqrt{1 + a^2} u, u_2 = v / \sqrt{1 + a^2},
\end{equation}
where indexes have been lowered for simplicity in the notation. The metric results to be
\begin{equation}
g = d u_1^2 + u_1^2 d u_2^2.
\end{equation}

In matrix form we get
\begin{equation}
g = \left( \begin{array}{ll}
     1 & 0\\
     0 & u_1^2
   \end{array} \right), g^{- 1} = \left( \begin{array}{ll}
     1 & 0\\
     0 & u_1^{- 2}
   \end{array} \right).
\end{equation}

The Christoffel symbols of the metric are defined for the Levi-Civita
connection as
\begin{equation}
\Gamma^i_{j k} = \frac{1}{2} g^{i l} \left( \frac{\partial g_{l
   k}}{\partial u_j} + \frac{\partial g_{l j}}{\partial u_k} - \frac{\partial
   g_{j k}}{\partial u_l} \right).
\end{equation}

We compute both components
\begin{equation}
\Gamma^1_{j k} = \frac{1}{2} g^{1 1} \left( \frac{\partial g_{1
   k}}{\partial u_j} + \frac{\partial g_{1 j}}{\partial u_k} - \frac{\partial
   g_{j k}}{\partial u_1} \right), 
\end{equation}
\begin{equation} \Gamma^2_{j k} = \frac{1}{2} g^{2 2} \left( \frac{\partial g_{2
   k}}{\partial u_j} + \frac{\partial g_{2 j}}{\partial u_k} - \frac{\partial
   g_{j k}}{\partial u_2} \right),
\end{equation}
and therefore
\begin{equation}
 \Gamma^1_{1 1} = 0, \Gamma^1_{2 2} = - \frac{1}{2} g^{1 1} \frac{\partial
   g_{2 2}}{\partial u_1} = - u_1, \Gamma^1_{12} = 0,
\end{equation}
\begin{equation}
\Gamma^2_{1 1} = 0, \Gamma^2_{2 2} = 0, \Gamma^2_{12} = \frac{1}{u_1}.
\end{equation}
The geodesic equations can be written in a very simple manner
\begin{equation}
\begin{array}{l}
     \ddot{u_1} - u_1 \dot{u_2}^2 = 0 \Rightarrow \ddot{u_1} -
     \dfrac{L^2}{u_1^3} = 0 \Rightarrow \ddot{u_1} \dot{u_1} -
     \dfrac{L^2}{u_1^3} \dot{u_1} = 0 \Rightarrow \dfrac{d}{d t} \left(
     \frac{1}{2} \dot{u}^2 + \dfrac{L^2}{2 u^2} \right) = 0,\\
     \ddot{u_2} + \dfrac{2}{u_1}  \dot{u_1} \dot{u_2} = 0 \Rightarrow
     \dfrac{d}{d t} (\ln \dot{u_2} u_1^2) = 0 \Rightarrow \dot{u_2} u_1^2 = L.
   \end{array}
\end{equation}
The parameters $L$ and $E$ are the angular momentum and the energy of the
particle. We express them using a more traditional notation, conventionally
used in physics, by replacing $u_{1}$ with a radial coordinate $r$ and $u_{2}$ with an angular coordinate $\phi$. With this replacement, we obtain
\begin{equation}
E = \frac{1}{2} (\dot{r}_0^2 + r_0^2 \dot{\phi}_0^2) = \frac{1}{2} v_0^2,
   \dot{\phi_0} r_0^2 = L.
\end{equation}
Thus the equations of motion are
\begin{equation}
\begin{array}{l}
     \dfrac{1}{2} \dot{r}^2 + \dfrac{L^2}{2 r^2} = E \Rightarrow \dfrac{d r}{d t}
     = \pm \sqrt{2 E - \dfrac{L^2}{r^2}}\\
     \dot{\phi} r^2 = L
   \end{array},
\end{equation}
where $v_{0}$ is the modulus of the speed of the particle.

We have two possible solutions: one towards the apex of the cone, and another going away from it. In terms of the initial conditions we have
\begin{equation}
\begin{array}{l}
     \dfrac{1}{2} \dot{r}^2 + \dfrac{L^2}{2 r^2} = E \Rightarrow \dfrac{d r}{d t}
     = \pm \sqrt{\dot{r_0}^2 + r_0^2 \dot{\phi}_0^2  \left( 1 -
     \dfrac{r_0^2}{r^2} \right)}\\
     \dot{\phi} r^2 = L
   \end{array}.
\end{equation}

Note that for the integral to be well-behaved, we must require
\begin{equation}
 \dot{r_0}^2 + \dot{\phi}_0^2 r_0^2 \left( 1 - \dfrac{r_0^2}{r^2} \right)
   \geqslant 0.
\end{equation}

If the trajectories go away from the apex of the cone, the previous equation is well behaved since $r>r_{0}$. On the contrary, we see that
\begin{equation}
r \geqslant \dfrac{r_0}{\sqrt{1 + \left( \dfrac{\dot{r_0}}{r_0 \dot{\phi}_0}
   \right)^2}} = \frac{r_0^2 \dot{\phi}_0}{v_0} \Rightarrow r \geqslant r_0
   \dfrac{v_{\phi_0}}{v_0}
\end{equation}

Thus if $v_{\phi_0} = 0$, again the equation is well behaved, but if $v_{\phi_0} = v_0 /2$, for example, we see that $r \geqslant r_0 / 2$. This means that the geodesics come back after going around the cone when they are launched towards the apex with some angular velocity, as shown in Fig.~\ref{fig:2}. We have now to integrate the differential equations
\begin{equation} 
\begin{array}{l}
     \dfrac{d r}{d t} = \pm \sqrt{2 E - \dfrac{L^2}{r^2}}\\
     \dot{\phi} = \dfrac{L}{r^2}
   \end{array}.
\end{equation}
The integration is straightforward, yielding the positive solutions
\begin{equation} 
\begin{array}{l}
     r (t) = \sqrt{\dfrac{L^2}{2 E} + 2 E (\beta + t)^2} \\
     \phi (t) = \phi_0 + \tmop{arc} \tan \left( \dfrac{2 E}{L} (t + \beta)
     \right) - \tmop{arc} \tan \left( \dfrac{2 E}{L} \beta \right)
   \end{array}.
\end{equation}
where we have introduced the parameter
\begin{equation} 
\begin{array}{l}
     \beta = \sqrt{\dfrac{r_0^2}{2 E} - \dfrac{L^2}{4 E^2}} \noplus =
     \dfrac{r_0}{v_0} \sqrt{1 - \left( \dfrac{v_{\phi_0}}{v_0} \right)^2},
   \end{array} \noplus
\end{equation}
for simplicity. Now we compute the velocities to later introduce the Riemann coordinates. We have the initial conditions
\begin{equation}
\begin{array}{l}
     \dot{r} (0) = v_{r_0} = c_1\\
     \dot{\phi} (0) =
     \dfrac{v_{\phi_0}}{r_0} = c_2 
   \end{array},
\end{equation}
where these initial conditions have been written by means of the constants $c_{1}$ and $c_{2}$. These two constants allow to rewrite the solutions in terms of the initial conditions and the time as
\begin{equation}
\begin{array}{l}
     r (t, c_1, c_2) = r_0  \sqrt{\dfrac{r_0^2 c_2^2}{c_1^2 + r_0^2 c_2^2
     \nocomma} + \left( \dfrac{c_1}{\sqrt{c_1^2 + r_0^2 c_2^2} \nocomma} +
     \sqrt{\dfrac{c_1^2}{r_0^2} + c_2^2} t \right)^2} \\
     \phi (t, c_1, c_2) = \phi_0 + \tmop{arc} \tan \left( \dfrac{c_1
     \nocomma}{r_0 c_2} + \dfrac{1}{c_2}  \left( \dfrac{c_1^2}{r_0^2} + c^2_2
     \right) t \right) - \tmop{arc} \tan \left( \dfrac{c_1 \nocomma}{r_0 c_2}
     \right)
   \end{array}.
\end{equation}
Considering the point $r_0 = 1$ and $\phi_0 = 0$ as the point where the origin of the Riemann coordinate system is placed, we derive
\begin{equation} 
\begin{array}{l}
     r (t, c_1, c_2) = \dfrac{1}{\sqrt{c_1^2 + c_2^2 \nocomma}}  \sqrt{c_2^2 +
     (c_1 + (c_1^2 + c_2^2) t)^2} \\
     \phi (t, c_1, c_2) = \tmop{arc} \tan \left( \dfrac{c_1 \nocomma}{c_2} +
     \dfrac{1}{c_2}  (c_1^2 + c^2_2) t \right) - \tmop{arc} \tan \left(
     \dfrac{c_1 \nocomma}{c_2} \right)
   \end{array}.
\end{equation}

A suitable reparametrization gives
\begin{equation} 
\begin{array}{l}
     r (k t, c_1, c_2) = \dfrac{1}{\sqrt{c_1^2 + c_2^2 \nocomma}}  \sqrt{c_2^2
     + (c_1 + (c_1^2 + c_2^2) k t)^2} \\
     \phi (k t, c_1, c_2) = \tmop{arc} \tan \left( \dfrac{c_1 \nocomma}{c_2} +
     \dfrac{1}{c_2}  (c_1^2 + c_2^2) k t \right) - \tmop{arc} \tan \left(
     \dfrac{c_1 \nocomma}{c_2} \right)
   \end{array},
\end{equation}
what sheds light into the properties of the mapping that will be used to define the Riemann coordinate system \cite{krey}. These properties are
\begin{equation} 
\begin{array}{l}
     r (t, k c_1, k c_2) = r (k t, c_1, c_2)\\
     \phi (t, k c_1, k c_2) = \phi (k t, c_1, c_2)
   \end{array} \Rightarrow \begin{array}{l},
     r (1, t c_1, t c_2) = r (t, c_1, c_2)\\
     \phi (1, t c_1, t c_2) = \phi (t, c_1, c_2)
   \end{array}
\end{equation}
and allow to define the Riemann coordinate system through the exponential mapping.

In summary, we can define the new coordinates $v_{\alpha} = c_{\alpha} t$ and arrive at the following coordinate transformation $\Phi_i (v_1, v_2) = u_i (1, v_1, v_2)$, which is explicitly written as
\begin{equation} 
\begin{array}{l}
     u_1 = \Phi_1 (v_1, v_2) = \sqrt{\dfrac{v_2^2 + (v_1 + (v_1^2 +
     v_2^2))^2}{v_1^2 + v_2^2}} = \sqrt{1 + 2 v_1 + v_1^2 + v_2^2} \\
     u_2 = \Phi_2 (v_1, v_2) = \tmop{arc} \tan \left( \dfrac{v_1}{v_2} +
     \dfrac{\nocomma v_1^2 + v_2^2}{v_2} \right) - \tmop{arc} \tan \left(
     \dfrac{v_1 \nocomma}{v_2} \right)
   \end{array},
\end{equation}
and can be further simplified by means of the formula for difference of the inverses of the tangent of two angles. The concluding result is
\begin{equation} 
\begin{array}{l}
     \Phi_1 (v_1, v_2) = \sqrt{1 + 2 v_1 + v_1^2 + v_2^2} \\
     \Phi_2 (v_1, v_2) = \tmop{arc} \tan \left( \dfrac{v_2}{1+v_1}\right)
   \end{array}.
\end{equation}
Now we compute the Jacobian as
\begin{equation} \begin{array}{l}
     \dfrac{\partial \Phi_1}{\partial v_1} = \dfrac{1 + v_1}{\sqrt{1 + 2 v_1 +
     v_1^2 + v_2^2}}, \dfrac{\partial \Phi_1}{\partial v_2} =
     \dfrac{v_2}{\sqrt{1 + 2 v_1 + v_1^2 + v_2^2}} \\
     \dfrac{\partial \Phi_2}{\partial v_1} = - \dfrac{v_2}{(1 + v_1)^2 + v_2^2},
     \dfrac{\partial \Phi_2}{\partial v_2} = \dfrac{1 + v_1}{(1 + v_1)^2 +
     v_2^2}.
   \end{array}
\end{equation}

Therefore, at the point $P(0,0)$ the Jacobian is the identity, as expected. Note that if we perform a linear transformation at the point of the cone we simply get a new coordinate system $\tilde{v}^i$, but the coordinate transformation connecting the new coordinates and the new Riemann coordinate system remain the same.

\section{Switching geodesics at the apex of a cone}

To examine the concepts developed in the first section for a conical point,
we parametrize a cone as in the previous section. Thus we focus on an
apparently much more simple case, which are the geodesic solutions for which
$L = 0$, centred at the apex of the cone
\begin{equation} \begin{array}{l}
     \dot{r}^2 = 2 E \Rightarrow r = r_0 + v_0 t\\
     \dot{\phi} r^2 = 0 \Rightarrow \phi = \phi_0
   \end{array} \Rightarrow \begin{array}{l}
     r (t) = v_0 t\\
     \phi (t) = \phi_0
   \end{array}.
\end{equation}

We are simply following straight lines with uniform motion. If we consider the apex of the cone and want to set a coordinate system there, which is allowable everywhere except for the apex, we find that these coordinates serve for the purpose. But now a problem arises, since the geodesics all have a constant coordinate
\begin{equation} 
\begin{array}{l}
     u_1(t) = v_0 t\\
     u_2(t) = u_{20}
   \end{array}.
\end{equation}

This coordinates are known as \tmem{geodesic polar coordinates}. In this
case we can introduce Riemann coordinates by considering
\begin{equation} 
v_1 = u_1 \cos u_2, v_2 = u_1 \sin u_2,
\end{equation}
and now we have $c_1 = \cos u_2$ and $c_2 = \sin u_2$. Thus we arrive at
\begin{equation} 
v_1 = v_0 c_1 t, v_2 = v_0 c_2 t,
\end{equation}
and therefore we can write
\begin{equation} u_1 = \sqrt{v_1^2 + v_2^2}, u_2 = \arctan \left( \frac{v_2}{v_1} \right).
\end{equation}

Thus we have $\Phi_1 (v_1, v_2) = \sqrt{v_1^2 + v_2^2}$ and $\Phi_2 (v_1, v_2)
= \arctan \left( \dfrac{v_2}{v_1} \right)$. If we now perform an orthogonal and special linear
transformation, we see that
\begin{equation} 
\begin{array}{l}
     \tilde{u}_1 = \sqrt{\tilde{v}_1^2 + \tilde{v}_2^2} = \sqrt{v_1^2 + v_2^2}
     = u_1\\
     \tilde{u}_2 = \arctan \left( \dfrac{ \tilde{v}_2}{\tilde{v}_1} \right) =
     \arctan \left( \dfrac{\sin \alpha \cos u_2 + \cos \alpha \sin u_2}{\cos
     \alpha \cos u_2 - \sin \alpha \sin u_2} \right)
   \end{array}.
\end{equation}

Consequently, $F$ is defined as
\begin{equation} 
\begin{array}{l}
     \tilde{u}_1 = F_1 (u_1, u_2) = u_1\\
     \tilde{u}_2 = F_2 (u_1, u_2) = u_2 + \alpha
   \end{array}.
\end{equation}

After this computation, we can obtain the Jacobian of $\Phi$ as
\begin{equation} 
\begin{array}{l}
     \dfrac{\partial \Phi_1}{\partial \tilde{v}_1} =
     \dfrac{\tilde{v}_1}{\sqrt{\tilde{v}_1^2 + \tilde{v}_2^2}} = \cos
     \tilde{u}_2, \dfrac{\partial \Phi_1}{\partial \tilde{v}_2} =
     \dfrac{\tilde{v}_2}{\sqrt{\tilde{v}_1^2 + \tilde{v}_2^2}} = \sin
     \tilde{u}_2\\
     \dfrac{\partial \Phi_2}{\partial \tilde{v}_1} =
     \dfrac{\tilde{v}_1}{\tilde{v}_1^2 + \tilde{v}_2^2} = - \tilde{u}_1 \sin
     \tilde{u}_2, \dfrac{\partial \Phi_1}{\partial \tilde{v}_2} =
     \dfrac{\tilde{v}_2}{\tilde{v}_1^2 + \tilde{v}_2^2} = \tilde{u}_1 \cos
     \tilde{u}_2
   \end{array},
\end{equation}
and once again, for the same reasons, we have
\begin{equation} 
\begin{array}{l}
     \dfrac{\partial \Phi_1}{\partial v_1} = \cos u_2, \dfrac{\partial
     \Phi_1}{\partial v_2} = \sin u_2\\
     \dfrac{\partial \Phi_2}{\partial v_1} = - u_1 \sin u_2, \dfrac{\partial
     \Phi_1}{\partial v_2} = u_1 \cos u_2
   \end{array}.
\end{equation}

The Jacobian matrix is therefore written as
\begin{equation} 
\tilde{J} = \left( \begin{array}{ll}
     \cos \tilde{u}_2  & - \tilde{u}_1 \sin \tilde{u}_2\\
     \sin \tilde{u}_2 & \tilde{u}_1 \cos \tilde{u}_2
   \end{array} \right) = \left( \begin{array}{ll}
     \cos (u_2 + \alpha) & - u_1 \sin (u_2 + \alpha)\\
     \sin (u_2 + \alpha) & u_1 \cos (u_2 + \alpha)
   \end{array} \right) \nocomma,
\end{equation}
whose inverse is
\begin{equation} J^{- 1} = \left( \begin{array}{cc}
     \cos u_2  & \sin u_2\\
     - u_1^{- 1} \sin u_2 & u_1^{- 1} \cos u_2
   \end{array} \right) \nocomma.
\end{equation}

Finally, we can obtain a relation between the tangent vectors
\begin{equation} \dot{\tilde{u}}^i = J^i_j L^j_k \hat{J}_l^k \dot{u}^l = \dot{u}^l.
\end{equation}

More specifically, we have the two identities
\begin{equation} 
\begin{array}{l}
     \dot{\tilde{u}}^1 = \dot{u}^1 = c_1\\
     \dot{\tilde{u}}^2 = \dot{u}^2 = 0
   \end{array},
\end{equation}
as desired, since we have simply rotated the vector, without changing its
velocity.

\section{Computing the tensors $\Omega^i_j$ and $\Gamma^i_{j k}$
}

Here we first demonstrate that the last term appearing in Eq.~(72) are the
Christoffel symbols of the manifold. We start with a transformation of
coordinates from the manifold coordinates to the Riemann coordinates
\begin{equation} 
u^i = \Phi^i (v_1, v_2).
\end{equation}
Differentiating twice this expression yields
\begin{equation} 
\ddot{u}^i = \dfrac{\partial^2 \Phi^i}{\partial v^j \partial v^k} \dot{v}^k
   \dot{v}^j + \dfrac{\partial \Phi^i}{\partial v^j} \ddot{v}^j.
\end{equation}

Since in the Riemann coordinate system the movement is uniform, the second
term of the right hand side vanishes. Considering that the tangent components between
both coordinate systems are related by the Jacobian we obtain
\begin{equation} 
\ddot{u}^i - \dfrac{\partial^2 \Phi^i}{\partial v^j \partial v^k}
   \dfrac{\partial \Psi^k}{\partial u^l} \dfrac{\partial \Psi^j}{\partial u^s} \dot{u}^l \dot{u}^s = 0.
\end{equation}

Therefore we have that
\begin{equation} 
\Gamma^i_{j k} = - \dfrac{\partial^2 \Phi^i}{\partial v^l \partial v^s}
   \dfrac{\partial \Psi^l}{\partial u^j} \frac{\partial \Psi^s}{\partial u^k}.
\end{equation}

Concerning the computation of the tensor $\Omega^i_k$ appearing in Eq.~(72)
\begin{equation} 
\Omega^i_k = \dfrac{\partial \Phi^i}{\partial v^j} \varepsilon_m^j
   \dfrac{\partial \Psi^m}{\partial u^k},
\end{equation}
we begin by recalling that
\begin{equation} 
(g^{-1})^i_k = \dfrac{\partial \Phi^i}{\partial v^l}
   \dfrac{\partial \Phi_k}{\partial v_l},
\end{equation} 
where $g^{-1}$ is here a $(1,1)$-tensor with entries equal to those of the inverse of the metric tensor and the indexes have been lowered to indicate that the matrix has been transposed. Multiplying the previous equation by
\begin{equation} 
\Omega^i_k (g^{-1})^k_l = \dfrac{\partial \Phi^i}{\partial v^j} \varepsilon_m^j
   \dfrac{\partial \Phi_l}{\partial u_m}.
   \label{eq:c7}
\end{equation}

Thus we have the relation between matrices $\Omega g^{-1}=J \varepsilon J^{t}$. But the Levi-Civita tensor can be related to the volume element. As it is well-known, the volume element under coordinate transformations provides the determinant of the metric $|g|$. Expressed by means of the wedge product the relation reads
\begin{equation} 
du^{1} \wedge du^{2} = \sqrt{|g|} dv^{1} \wedge dv^{2}.
\end{equation}

Written in matrix form, we then have that $J^{t} \varepsilon J = \sqrt{|g|} \varepsilon$. Multiplying this equality by $J$ on the left side, by $J^{t}$ on the right, and recalling that $g^{-1}=J J^{t}$ and that $g$ is a symmetric matrix, we have that $J \varepsilon J^{t} = \sqrt{|g^{-1}|} \varepsilon$. Substitution in Eq.~\eqref{eq:c7} yields the desired result
\begin{equation} 
\Omega_k^i=\dfrac{1}{\sqrt{|g|}} \varepsilon^i_j g^j_k,
\end{equation}
where $g^j_k$ is again a $(1,1)$-tensor whose entries are equal to those of the metric tensor.

\end{appendices}

\end{document}